\documentclass[aps,prb,reprint,amsmath,amssymb,superscriptaddress,floatfix,notitlepage]{revtex4-2}
\usepackage{graphicx}%
\usepackage[colorlinks, citecolor=blue, linkcolor=blue]{hyperref}
\usepackage{xcolor}
\usepackage{gensymb}
\usepackage{float}
\usepackage{ragged2e} 
\usepackage{subcaption}   
\DeclareCaptionJustification{justified}{\justifying}
\newcommand{\KIEL}{Institute of Theoretical Physics and Astrophysics, University of Kiel, Leibnizstrasse 15, 24098 Kiel, Germany}
\newcommand{\KINSIS}{Kiel Nano, Surface, and Interface Science (KiNSIS), University of Kiel, 24118 Kiel, Germany}

\newcommand{\mmdot}[2]{\mathbf{m}_{#1}\cdot\mathbf{m}_{#2}}
\newcommand{\mmdotTB}[4]{\mathbf{m}_{#1}^{\text{#3}}\cdot\mathbf{m}_{#2}^{\text{#4}}}

\newcommand{\mvec}[1]{\mathbf{m}_{#1}}

\begin{document}

\title{Bilayer triple-Q state driven by interlayer higher-order exchange interactions}

\author{Bjarne Beyer}
\thanks{These two authors contributed equally to this work.}
\email[Contact author: ]{beyer@physik.uni-kiel.de}
\affiliation{\KIEL}

\author{Mara Gutzeit}
\thanks{These two authors contributed equally to this work.}
\affiliation{\KIEL}

\author{Tim Drevelow}
\affiliation{\KIEL}

\author{Isabel Schwermer}\affiliation{\KIEL}

\author{Soumyajyoti Haldar}
\email[Contact author: ]{haldar@physik.uni-kiel.de}
\affiliation{\KIEL}

\author{Stefan Heinze}
\email[Contact author: ]{heinze@physik.uni-kiel.de}
\affiliation{\KIEL}\affiliation{\KINSIS}

\date{\today}

\begin{abstract}
Using first-principles calculations and an atomistic spin model we predict the stabilization of a bilayer triple-Q state in an atomic Mn bilayer on Ir(111) due to 
interlayer higher-order exchange interactions.
Based on density functional theory (DFT) we study the magnetic
interactions and ground state in a Mn monolayer and bilayer on the Ir(111) surface. We calculate
the energy dispersion of spin spirals (single-Q states) to scan a large part of the magnetic phase space
and to obtain constants of pair-wise exchange interactions. 
By including spin-orbit coupling we determine the strength of the Dzyaloshinskii-Moriya interaction.
To reveal the role of higher-order exchange interactions in these films, we consider multi-Q states 
obtained by a superposition of spin spirals. For the Mn monolayer in fcc stacking on Ir(111), the 
triple-Q state exhibits the lowest total energy in DFT, while the N\'eel state is most favorable for hcp stacking. 
For the Mn bilayer on Ir(111), two types of the triple-Q state are possible. In both magnetic
configurations, a triple-Q
state occurs within each of the Mn layers. However, only in one of them the spin alignment between the
layers is such that nearest-neighbor spins of different layers also exhibit the tetrahedron angles which
characterize the triple-Q state. We denote this state -- which has the lowest total energy in our
DFT calculations -- as the ideal bilayer triple-Q state. This state exhibits significant topological
orbital moments within each of the two Mn layers which are aligned in parallel resulting in
a large topological orbital magnetization.
We interpret the DFT results within an atomistic spin model which includes pair-wise Heisenberg
exchange, the Dzyaloshinskii-Moriya interaction, as well as higher-order exchange interactions. 
We classify the different types of higher-order interactions into intralayer terms, i.e.~acting
only between spins within one of the layers, and interlayer interactions, in which spins in both layers are involved.
We point out the role of an odd or even distribution of spins in the multi-spin
interactions between the layers.
Finally, we demonstrate that the stabilization of the ideal bilayer triple-Q state in a Mn bilayer on Ir(111)
can only be explained upon taking the effect of interlayer higher-order exchange interactions into
account.
\end{abstract}

\maketitle

\section{Introduction}
Non-collinear spin structures have recently moved into the focus of research in spintronics and
nanomagnetism due to their promise for future applications in logic, memory, or novel computing
devices \cite{Fert2017,Back2020,Rimmler2024}. The magnetic properties of a material are often described within
an atomistic
spin model which 
contains the Heisenberg pair-wise exchange interaction, the Dzyaloshinskii-Moriya 
interaction \cite{Moriya1960,Dzyaloshinskii1957}
as well as the magneocrystalline anisotropy energy. However, in the last decade, compelling evidence 
has been given for the importance of higher-order exchange interactions 
\cite{Takahashi_1977,MacDonald88,Hoffmann2020}
in ultrathin transition-metal 
films by combined experimental and theoretical studies. In these studies a number of intriguing magnetic 
ground states have been identified such as the triple-Q state 
\cite{Kurz2001,Spethmann2020,Haldar2021,nickel2023coupling},
conical spin spirals \cite{Yoshida2012,Weber2023}, the $uudd$ state~\cite{Kroenlein2018,Romming2018},
and nanoskyrmion lattices 
\cite{Heinze2011,vonBergmann2015,Gutzeit2023,Nickel2024}. Higher-order interactions can also play an
important role for the stabilization of topological spin structures such as isolated
magnetic skyrmions or antiskyrmions \cite{Paul2020}. 

So far, mainly ultrathin films consisting of a single 
atomic magnetic layer have been considered. For such systems a spin model containing higher-order
exchange interactions has been developed \cite{Hoffmann2020}. However, spin models including the
effect of higher-order interactions between layers in magnetic bilayers or multilayers are missing.
An understanding of magnetic ground states and higher-order interactions in bilayers is non-trivial 
as interactions with vertical components and truly three-dimensional hopping processes come into play.

In this work, we apply density functional theory (DFT) to explore the complex magnetism of both a hexagonal Mn monolayer (ML) and a hexagonal Mn bilayer (BL) grown pseudomorphically on the Ir(111) surface and directly compare the results to a freestanding Mn BL. We perform total energy DFT calculations for various collinear and 
non-collinear magnetic states with and without including spin-orbit coupling. We find strong antiferromagnetic (AFM) exchange
interactions for the Mn ML and a stacking dependent magnetic ground state. For hcp Mn stacking,
we obtain the N\'eel state with 120$^\circ$ between adjacent spins  as the ground state consistent with recent experimental \cite{Rodriguez2024} and theoretical \cite{aldarawsheh2023intrinsic} work.
For fcc Mn stacking, we predict a triple-Q state with tetrahedron angles between adjacent spin moments. 

For the Mn BL on Ir(111), we 
obtain AFM exchange interactions within the top and bottom Mn layer as well as AFM exchange
between the two layers. As a result a row-wise AFM state with an effective AFM coupling 
between the layers is most favorable among all spin spiral states. However, upon considering
superposition states of spin spirals, i.e.~multi-Q states, we find that a triple-Q state
is the energetically lowest magnetic configuration in the bilayer. For a magnetic bilayer
two types of the triple-Q state are possible. In both of them the spins of adjacent Mn atoms
within a layer exhibit tetrahedron angles. However,
only one of the two shows
tetrahedron angles between adjacent spins of different layers. This ideal bilayer triple-Q
state
is favored
for the Mn BL on Ir(111). For this state the spin moments exhibit an effective 
AFM coupling between the layers and the total spin moment vanishes. However, 
significant topological orbital moments are induced in the triple-Q state
due to its non-coplanarity
\cite{hanke2016role,Haldar2021,nickel2023coupling}.
In the ideal 
bilayer triple-Q state, the topological orbital moments of the two layers couple 
ferromagnetically resulting in a significant total topological orbital magnetization \cite{Saxena2024}.

The core of our computational DFT approach is the calculation of the energy dispersion of
flat spin spirals propagating either in only one or simultaneously in two magnetic layers.
By mapping the total energies to a Heisenberg model, we can determine
both intra- and interlayer pair-wise exchange constants.
While the 
treatment of spin spiral states in magnetic MLs represents a well-established method to extract the intralayer Heisenberg exchange parameters, the calculation of the
spin spiral
energy dispersion of two interacting magnetic layers is
a computationally and conceptually much more demanding task.
We present a detailed introduction to the underlying
framework for the model of pair-wise Heisenberg exchange in hexagonal magnetic bilayers. 

We transfer the concept of multi-Q 
states -- such as the up-up-down-down 
($uudd$) \cite{Hardrat2009} and triple-Q \cite{Kurz2001} state -- from magnetic MLs to BLs. Similar to their ML analogues, the prototypical multi-Q states of hexagonal BLs are energetically strictly degenerate with their corresponding 1Q building blocks regarding the pairwise Heisenberg exchange. We develop an atomistic spin model which includes not only
intralayer HOI \cite{Hoffmann2020} but also interlayer HOI. By classifying the HOIs into even and odd terms
with respect to the distribution of spins
in the top and bottom layer
we demonstrate that the ideal bilayer 
triple-Q (3Q) state in the
Mn BL on Ir(111) is stabilized by the interplay of AFM interlayer pair-wise exchange
and interlayer higher-order exchange.

The paper is organized in the following way. Section~\ref{sec:Heisenberg_HOI_theory} serves as an introduction to the calculation methodology of spin spiral (1Q) and multi-Q states for magnetic mono- and bilayers from first-principles. We discuss how the obtained
total energies can be mapped to an atomistic spin model  
which includes 
both intra- and interlayer pairwise Heisenberg and higher-order exchange interactions.
In section~\ref{sec:Comp_methods} the applied computational DFT methods are described. The main part of the paper (section~\ref{sec:Results}) presents DFT total energy calculations of collinear and non-collinear
spin states for
Mn MLs on Ir(111), a freestanding Mn bilayer, and Mn BLs on Ir(111) as well as interpretation of the data within the context of the atomistic spin model developed in our work including intra- and interlayer HOI. The key aspects and results of the present work are summarized in section~\ref{sec:Conclusion}.  

\section{Atomistic spin model}
\label{sec:Heisenberg_HOI_theory}

In order to interpret the total energy DFT calculations of different collinear and non-collinear magnetic
states including spin-orbit coupling, we apply an atomistic spin model that is based on the interactions
between normalized magnetic moments $\mathbf{m}_i$ localized at atoms at
lattice sites $i$. 
The general form of the Hamiltonian is given by
\begin{equation}
	\label{eq:H_general}
	\begin{split}
		H=&
		-\sum_{ij}J_{ij}(\mathbf{m}_i\cdot\mathbf{m}_j) 
		-\sum_{ij}\mathbf{D}_{ij}\cdot(\mathbf{m}_i\times\mathbf{m}_j)\\
		&-\sum_{ijkl}K_{ijkl}(\mathbf{m}_i\cdot\mathbf{m}_j) (\mathbf{m}_k\cdot\mathbf{m}_l) 
		-\sum_{i} K_{\rm u} (m^z_i)^2\\
	\end{split}
\end{equation}
and includes pair-wise Heisenberg exchange, the Dzyaloshinskii-Moria interaction (DMI) arising from spin-orbit coupling (SOC) \cite{Dzyaloshinskii1957,Moriya1960}, higher-order exchange 
interactions \cite{Takahashi_1977,MacDonald88,Hoffmann2020}, and the
uniaxial magnetocrystalline anisotropy.
The strength of these terms is given
by the respective constants $J_{ij}$, $\mathbf{D}_{ij}$, $K_{ijkl}$, and $K_{\rm u}$. The magnitude of the magnetic moment 
at site $i$ is denoted by $M_i$ and the magnetic moment is 
given by $\mathbf{M}_i = M_i \mathbf{m}_i$.

To parameterize the model, the total energies of different collinear and non-collinear
magnetic states are computed by DFT with and without including spin-orbit coupling
(details in Sec.~\ref{sec:Comp_methods}) and fitted to the spin model by minimizing the residual sum of squares. 
Depending on the geometry of the system and the interactions that need to be calculated, 
different magnetic states must be considered within DFT.
In the next two subsections we describe how this
spin model is applied to the cases of magnetic mono- and bilayers.

\subsection{Magnetic monolayers}
\label{subsec:ML_dispersion}
The largest contribution to the energy of a magnetic state usually stems from the first term of Eq.~(\ref{eq:H_general}), the pair-wise Heisenberg exchange,
emerging from the interplay of Pauli principle and  Coulomb interaction.
The sign of the exchange constant $J_{ij}$ 
determines whether the system at hand prefers a parallel orientation, i.e.~a ferromagnetic (FM) coupling ($J_{ij}>0$), or an antiparallel alignment, i.e.~an antiferromagnetic (AFM) coupling ($J_{ij}<0$), between two magnetic
moments $\mathbf{m}_i$ and $\mathbf{m}_j$. However, often
the situation does not turn out to be that trivial and bilinear exchange beyond nearest neighbors with different signs of $J_{ij}$ needs to be taken into account resulting in the formation of non-collinear magnetic ground states such as spin spirals on account of exchange frustration.

\begin{figure*}[htb]
	\centering
	\includegraphics[width=0.9\textwidth]{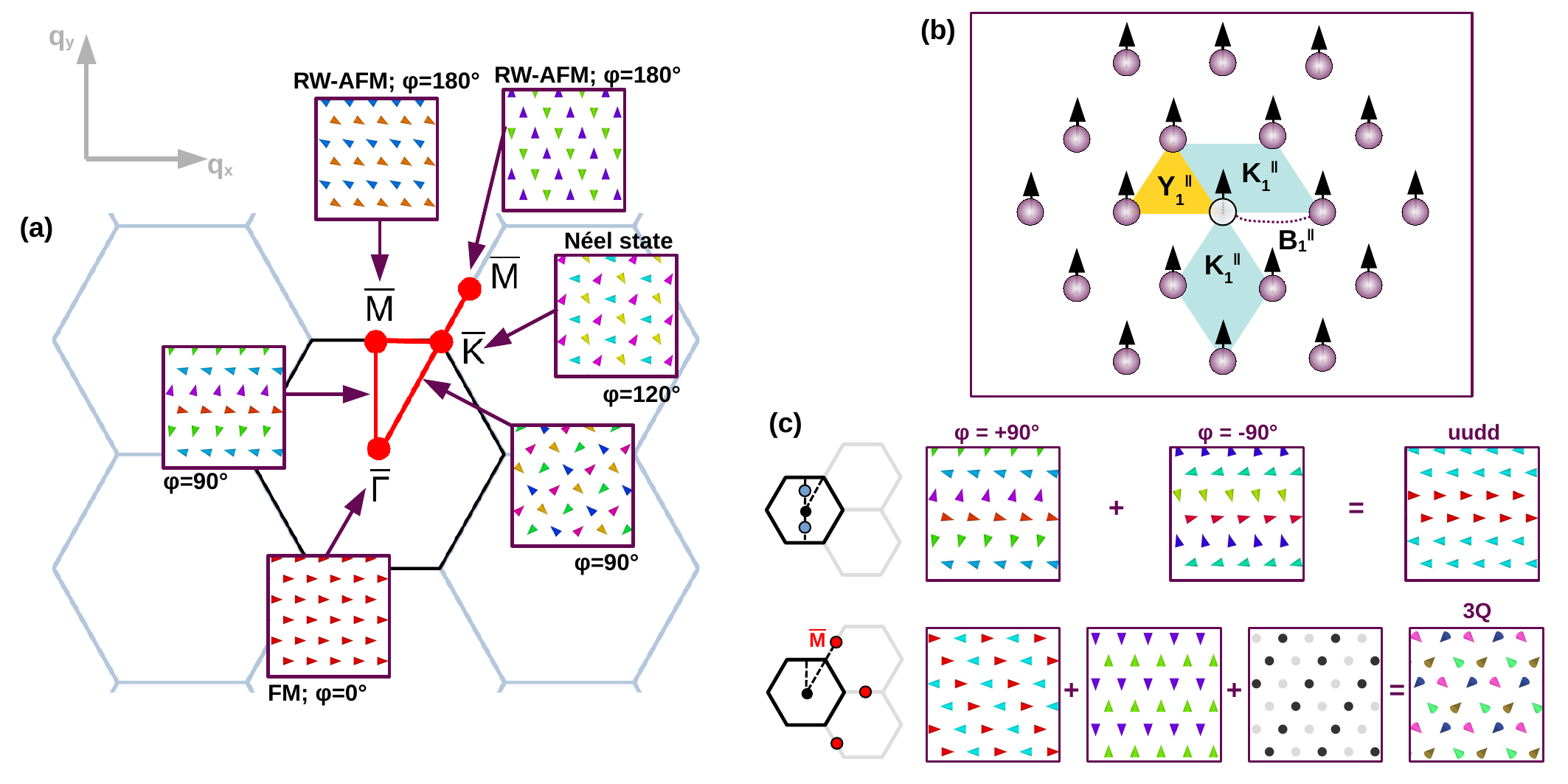}
	\caption{Sketch of the 2D hexagonal Brillouin zone (BZ) with a selection of possible flat spin spiral states, illustration of intralayer higher-order exchange interactions (HOIs) on the 2D hexagonal lattice, and examples for the construction of multi-Q states for a magnetic ML.
		(a) Irreducible part of the hexagonal 2D-BZ (encircled in red) with its high symmetry paths as well as the spin structures arising at the high symmetry points $\overline{\Gamma}$, $\overline{\text{K}}$ and $\overline{\text{M}}$.
		(b) Minimal hopping paths within the NN approximation for an electron taking part in the biquadratic interaction denoted by its respective strength $B_1^{\parallel}$, the 4-spin 3-site ($Y_1^{\parallel}$) and the 4-spin 4-site interaction ($K_1^{\parallel}$). For the latter, two out of 12 possible diamonds are depicted in light blue color and one out of six possible triangles for $Y_1^{\parallel}$ is illustrated in orange color. (c) Formation of a collinear $uudd$ state from a superposition of two counterpropagating 90$^{\circ}$ spin spirals at $\mathbf{q}$=$\pm \frac{1}{2}\overline{\Gamma\text{M}}$ (in a similar fashion, the other $uudd$ state is constructed from a superposition of left- and right-rotating 90$^{\circ}$ spin spirals at $\mathbf{q}$=$\pm \frac{3}{4}\overline{\Gamma\text{K}}$) and the non-collinear 3Q state originating from the superposition of three $\mathbf{q}$-vectors at the symmetry-equivalent $\overline{\text{M}}$-points.    } 
	\label{fig:ML_theory}
\end{figure*} 

The second term in Eq.~(\ref{eq:H_general}) describes the antisymmetric part of the exchange interaction, the DMI. Necessary ingredients for this relativistic effect to occur are both SOC and an inversion-asymmetric environment. In contrast to the symmetric pairwise Heisenberg exchange, the DMI prefers canted spin structures with a well-defined rotational sense owing to the anticommutativity of the cross product between two involved magnetic moments.  

Due to the high symmetry of a single magnetic hexagonal layer, 
interactions between two atoms with the same distance always share the same interaction constant.
Therefore, the exchange constants $J_{ij}$ as well as the magnitudes of the DM vectors $D_{ij}$
can be organized in shells based on the distance of the involved atoms $i$ and $j$
(see Appendix~\ref{sec:fittingfunctions}).
By ordering the shells with increasing distance, each shell is given a number $\sigma\geq 1$ 
that uniquely identifies the interaction constants $J_\sigma$ and $D_{\sigma}$.

Flat spin spirals
with varying spiral vector $\mathbf{q}$ along high-symmetry directions
of the Brillouin zone
provide a set of magnetic states that make it possible to determine $J_\sigma$ and $D_{\sigma}$
by means of their energy dispersion E($\mathbf{q}$) obtained via DFT.
For the case of the pairwise Heisenberg exchange, the individual contributions of each interaction constant $J_\sigma$ to E($\mathbf{q}$) 
are given in Appendix~\ref{sec:fittingfunctions}.
At first, all $J_\sigma$ are calculated by computing the energy of spin spirals without SOC since DMI arises from the latter. 
After that, the contribution of SOC on the energy is determined for each spin spiral state and mapped onto the DMI in the atomistic spin  model.

Because of the periodicity of the atomic lattice only $\mathbf{q}$ vectors 
in the first two-dimensional (2D) Brillouin zone (BZ) need to be taken into account for the calculation of E($\mathbf{q}$).
As shown in Fig.~\ref{fig:ML_theory}(a), our DFT calculations are restricted to spin spiral vectors along the high symmetry paths $\overline{\Gamma\text{M}}$ and $\overline{\Gamma\text{K}\text{M}}$ 
around the irreducible wedge of the hexagonal BZ. In a flat spin spiral, the direction of a magnetic moment at a specific lattice site $\textbf{R}_i$ is given by
$\textbf{m}_i$=$(\cos(\mathbf{q}\cdot \mathbf{R}_i),\sin(\mathbf{q}\cdot \mathbf{R}_i),0)^T$ where the dot product of the reciprocal space vector $\mathbf{q}$
and the real space vector $\textbf{R}_i$
yields the angle $\varphi_i$, denoting the spin orientation in the $xy$-plane.
Note, that the rotation plane of the spin
spiral can be chosen arbitrarily as along as
SOC is neglected.

Fig.~\ref{fig:ML_theory}(a) illustrates
important spin spiral states of a hexagonal magnetic ML arising at the high symmetry points of the 2D BZ: at the $\overline{\Gamma}$-point corresponding to $\mathbf{q}=0$ the FM state with a relative angle of
$\varphi =0^{\circ}$
between adjacent moments occurs, whereas at the $\overline{\text{M}}$-point the row-wise antiferromagnetic (RW-AFM)
($\varphi =180^{\circ}$)
and at the $\overline{\text{K}}$-point the N\'eel state with angles of 120$^{\circ}$ between adjacent moments can be found. Note that due to the symmetry of the pairwise Heisenberg exchange (see first term of Eq.~(\ref{eq:H_general})) the energies of left-($-\mathbf{q}$) and right-rotating ($+\mathbf{q}$) spin spirals with the same angle $\varphi$ between neighboring spins are degenerate if SOC and thereby DMI is neglected.

Spin spiral states also denoted as single-Q (1Q) states with symmetry-equivalent $\mathbf{q}$ vectors can be coupled to so-called multi-Q states by higher-order interactions (HOI) beyond the model of pair-wise Heisenberg exchange arising in fourth order perturbation theory from a multi-band Hubbard model~\cite{Hoffmann2020} (third term in Eq.~(\ref{eq:H_general})). For a magnetic ML, we restrict the discussion to the following 
HOI terms 
\cite{Hoffmann2020} throughout this paper: 

\begin{align}
	\label{eq:H_HOI_par}
	\notag H_{\text{HOI}}^{\parallel} =   -\sum_{ \langle ij \rangle} B_1^{\parallel}&( \mvec i \cdot \mvec j)^2\\
	\qquad \qquad -2 \sum_{ \langle ijk \rangle}Y_1^{\parallel}&(\mmdot ij )(\mmdot ik)\\
	\notag -\sum_{ijkl}K_1^{\parallel}\big[&(\mmdot ij)(\mmdot kl)\\[-10pt]
	\notag     +&(\mmdot il)(\mmdot jk)\\
	\notag     -&(\mmdot ik)(\mmdot jl)\big]
\end{align}

The coupling strengths within the magnetic layer are denoted by $B_1^{\parallel}$, $Y_1^{\parallel}$ and $K_1^{\parallel}$ for the biquadratic,
4-spin 3-site, and
4-spin 4-site interaction,
respectively. In a similar way, the intralayer pairwise Heisenberg exchange constants $J_\sigma$ and the magnitudes of the DMI vectors $D_{\sigma}$ will obtain the superscript $\parallel$
, denoting an interaction between moments within one layer,
in this work. Fig.~\ref{fig:ML_theory}(b) 
illustrates the interacting moments of minimal distances of the above-mentioned interactions: For the 4-spin 3-site (4-site) interaction the moments lie on an equilateral triangle (diamond).
The interacting moments are nearest neighbors (NN), except for the 4-spin 4-site interaction on the diamond, which 
contains a next-nearest neighbor (NNN) connection leading to the minus sign in the last term of 
Eq.~(\ref{eq:H_HOI_par}), i.e.~if the NNN spin moments appear in one of the two 
scalar products \cite{Hoffmann2020}.

On the hexagonal lattice of a magnetic ML, three prototypical multi-Q states can arise from the superposition of spin spiral (1Q) states due to HOI. These are two collinear up-up-down-down ($uudd$) states~\cite{Hardrat2009} generated from a superposition of two 90$^{\circ}$ spin spirals with opposite rotational sense (at $\mathbf{q}$=$\pm \frac{1}{2}\overline{\Gamma\text{M}}$ and $\mathbf{q}$=$\pm \frac{3}{4}\overline{\Gamma\text{K}}$) and one three-dimensional non-collinear 
3Q state \cite{Kurz2001,Spethmann2020}
originating from a superposition of three $\mathbf{q}$-vectors at symmetry-equivalent $\overline{\text{M}}$-points (see Fig.~\ref{fig:ML_theory}(c)). If only pairwise Heisenberg exchange was present, the multi-Q states and their respective single-Q states would be energetically degenerate. However,
the HOI interactions introduced above lift this degeneracy according to the following set of coupled equations~\cite{Hoffmann2020}:

\begin{equation}
	\label{eq:HOI_energy1}
	\Delta E_{\overline{\text{M}}} = E_{\overline{\text{M}}}^{\text{3Q}}-E_{\overline{\text{M}}}^{\text{1Q}} = \frac{16}{3}(2K_1^{\parallel} + B_1^{\parallel} - Y_1^{\parallel})
\end{equation}
\begin{equation}
	\label{eq:HOI_energy2}
	\Delta E_{\frac{1}{2}\overline{\Gamma\text{M}}}= E^{\text{uudd}}_{\frac{1}{2}\overline{\Gamma\text{M}}} - E_{\frac{1}{2}\overline{\Gamma\text{M}}}^{\text{1Q}} = 4(2K_1^{\parallel} - B_1^{\parallel} - Y_1^{\parallel})
\end{equation}
\begin{equation}
	\label{eq:HOI_energy3}
	\Delta E_{\frac{3}{4}\overline{\Gamma\text{K}}}= E^{\text{uudd}}_{\frac{3}{4}
		\overline{\Gamma\text{K}}} - E_{\frac{3}{4}
		\overline{\Gamma\text{K}}}^{\text{1Q}} = 4(2K_1^{\parallel} - B_1^{\parallel} + Y_1^{\parallel})
\end{equation}

Here, $E^{\text{uudd}}_{\frac{1}{2}\overline{\Gamma\text{M}}}$, 
$E^{\text{uudd}}_{\frac{3}{4}\overline{\Gamma\text{K}}}$ and 
$E_{\overline{\text{M}}}^{\text{3Q}}$ denote the total energies of the $uudd$ states along $\overline{\Gamma {\rm M}}$ and
$\overline{\Gamma {\rm K}}$
and the 3Q state (cf.~Fig.~\ref{fig:ML_theory}(c)), respectively, whereas 
$E_{\frac{1}{2}\overline{\Gamma\text{M}}}^{\text{1Q}}$,
$E_{\frac{3}{4}
	\overline{\Gamma\text{K}}}^{\text{1Q}}$,
and 
$E_{\overline{\text{M}}}^{\text{1Q}}$
are the total energies of the corresponding 1Q states.

\subsection{Magnetic bilayers}
\label{subsec:BL_dispersion}

\begin{figure}[htb]
	\centering
	\includegraphics[width=1\linewidth]{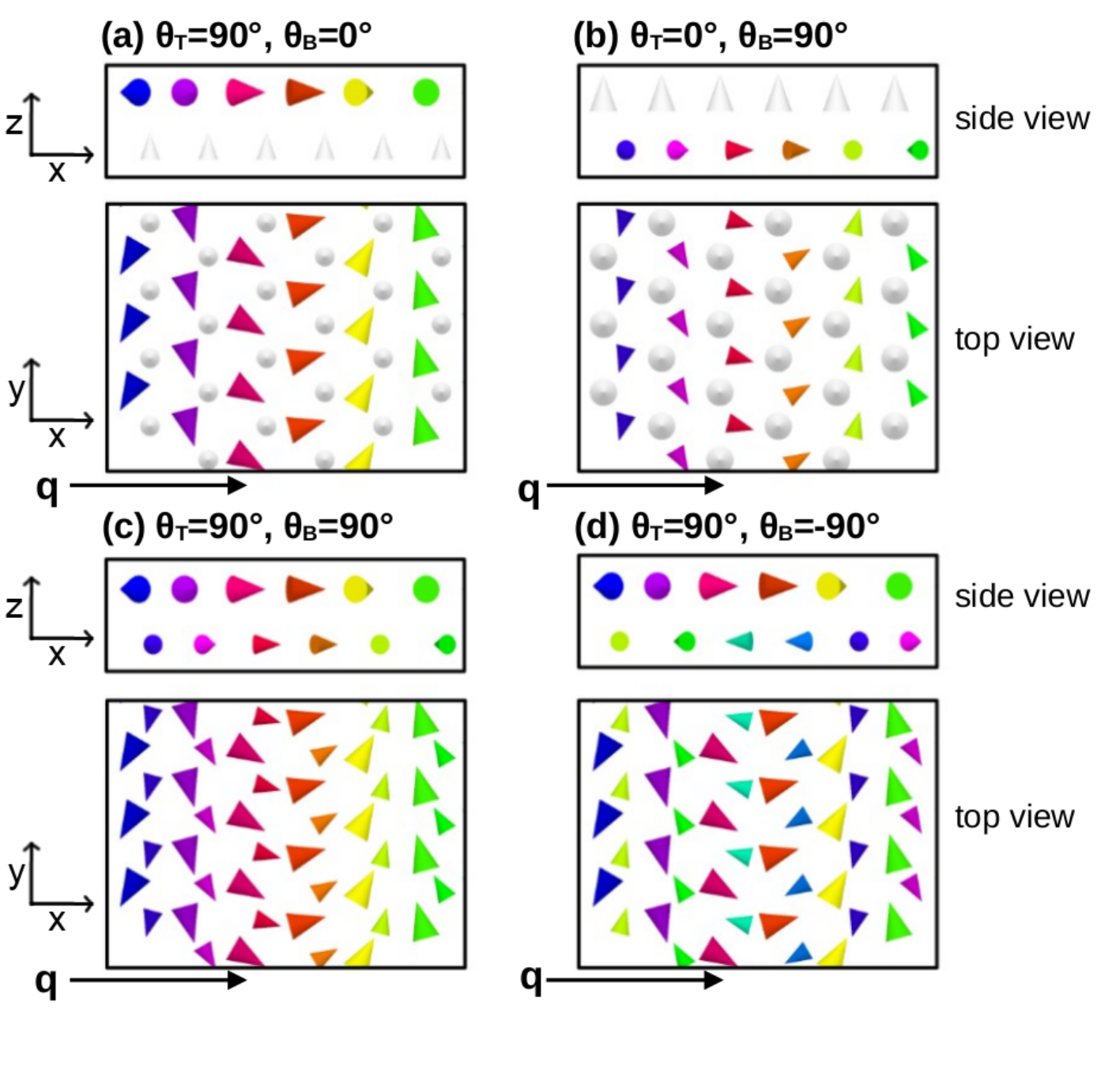}
	\caption{
		Spin spiral setups used for the DFT calculation of pair-wise Heisenberg intra- and interlayer
		exchange constants 
		in atomic hexagonally arranged magnetic bilayers.
		Magnetic moments belonging to the top layer are displayed larger compared to the bottom layer magnetic moments.
		(a,b) Spin spiral propagating along the wave vector $\mathbf{q}$
		either only in the top or the bottom layer, respectively,
		while the magnetic moments of the other
		layer are oriented perpendicular to the rotation plane of the spin spiral. 
		(c,d) Spin spirals propagating simultaneously in both magnetic layers with parallel and antiparallel
		alignment of spins between the layers, respectively.
		For every magnetic configuration the upper (lower) image illustrates its respective side (top) view.
		The magnetic configurations can be described by Eq.~(\ref{eq:spin_spiral}) and the angles 
		$\theta_{\rm T}$ and $\theta_{\rm B}$
		are
		given above each panel.
	}
	\label{fig:BL_SS_setup}
\end{figure}

With the introduction of a second magnetic layer, 
the distance of two atoms is no longer enough to determine if interactions share the same interaction constant since it becomes important which layer (top or bottom) the two interacting atoms belong to.
The four possible interactions top-top, top-bottom, bottom-top and bottom-bottom
each share their own sets $J_\sigma^\text{T}$, $J_\sigma^{\text{TB}}$, $J_\sigma^{\text{BT}}$, and $J_\sigma^\text{B}$,
respectively. 

These considerations lead
us to the pair-wise exchange Hamiltonian of a magnetic bilayer, describing interactions of two magnetic moments $\mvec i^{\rm T}$ and $\mvec j^{\rm B}$ lying in the top (T) or bottom (B) layer:
\begin{equation}
	\begin{split}
		H_{\rm BL} =  &-\sum\limits_{i\neq j} J_{ij}^{\text{T}} ( \mmdotTB ijTT) 
		- \sum\limits_{i \neq j} J_{ij}^{\text{B}} ( \mmdotTB ijBB)\\
		&-\sum\limits_{ij} J_{ij}^{\text{TB}} ( \mmdotTB ijTB) - \sum\limits_{ij} J_{ij}^{\text{BT}} ( \mmdotTB ijBT)
		\label{eq:H_BL}
	\end{split}
\end{equation}

Here, the first two contributions describe the interaction of spins within the individual magnetic layers, i.e. similar to a magnetic ML, and are hence referred to as intralayer exchange, while the latter two denote the interlayer exchange between the top and bottom magnetic layer and vice versa. As the corresponding exchange parameters are symmetric and the dot product of the moments is commutative, i.e. $\mathbf{m}_i^{\text{T}} \cdot \mathbf{m}_j^{\text{B}}=\mathbf{m}_i^{\text{B}} \cdot \mathbf{m}_j^{\text{T}}$, the two %
terms yield the same value thereby sharing the same exchange constant $J_\sigma^{\text{TB}}=J_\sigma^{\text{BT}}$
denoted $J_\sigma^\perp$ in the following.
For a symmetric magnetic bilayer, e.g.~a freestanding Mn bilayer
considered as a reference system in our DFT calculations presented
below, both magnetic 
layers are equivalent. In that case
the exchange constants of top and bottom layer coincide and
are denoted for the shells 
as $J_{\sigma}^{\parallel}=J_{\sigma}^{\text{T}}=J_{\sigma}^{\text{B}}$.

Hence, one can rewrite the Hamiltonian for the symmetric bilayer as
\begin{equation}
	\begin{split}
		H_{\text{s-BL}} =  -\sum\limits_{i\neq j, L} J_{ij}^{\parallel} 
		\left( \mathbf{m}_i^L \cdot \mathbf{m}_j^L   \right)
		-\sum\limits_{ij, L \ne L'} J_{ij}^{\perp} 
		\left( \mathbf{m}_i^L \cdot \mathbf{m}_j^{L'}   \right)
		\label{eq:H_BL_sym}
	\end{split}
\end{equation}

with the layer index $L \in \{ \text{T}, \text{B} \}$.
The pair-wise intra- and interlayer Heisenberg exchange constants 
can be determined separately via DFT 
calculations by taking different spin spiral setups into account. In case of two magnetic layers, the definition of the direction of a magnetic moment in a spin spiral state as introduced in the previous section needs to be adapted according to 
\begin{equation}
	\mathbf{m}_i^L=(\cos(\varphi_i^L)\sin(\theta_L),\sin(\varphi_i^L)\sin(\theta_L),\cos(\theta_L))^T   
	\label{eq:spin_spiral}
\end{equation}
for a spin at site $i$ in layer $L \in \{ \text{T},\text{B} \}$ and $\varphi_i^L=\mathbf{q} \cdot \mathbf{R}_i^L$.
Only the cone angle $\theta_L$ which describes the canting of the spins with respect to their rotation axis in layer $L$
varies between the two magnetic layers. 

For $\theta_\text{T}=90^{\circ}$ and $\theta_\text{B}=0^{\circ}$ (see Fig.~\ref{fig:BL_SS_setup}(a)) the flat spin spiral propagates only in the top magnetic layer while the moments of the bottom layer are aligned perpendicular to the overlying non-collinear spin structure thus making the interlayer Heisenberg exchange for the whole setup vanish. As  the intralayer exchange of the bottom layer remains constant, the only term left is the intralayer exchange of the top magnetic layer. With the same reasoning, the intralayer exchange constants of the bottom layer are obtained by letting the spin spiral propagate solely in this layer (Fig.~\ref{fig:BL_SS_setup}(b)). These types of DFT calculations can be restricted to $\mathbf{q}$ vectors along the high symmetry directions of the 2D hexagonal BZ as shown in Fig.~\ref{fig:ML_theory}(a) since the distances between neighboring moments in the individual layers correspond to those of the hexagonal magnetic ML. Consequently, the symmetry zone of spin spirals for the computation of the intralayer Heisenberg exchange is the usual hexagonal 2D BZ.

For the case that a flat spin spiral propagates simultaneously in both magnetic layers, i.e.~$\theta_\text{T}=90^{\circ}$ and $\theta_\text{B}=\pm 90^{\circ}$, 
two alignments between the spins of the layers need to be taken into account for the bilayer magnetisation.
In the first case, $\theta_\text{B}=90^{\circ}$, the same magnetisation is present in the two layers, neglecting the phase-shift due to the hexagonal stacking. The second case introduces a spin flip of the bottom layer, through $\theta_\text{B}=-90^{\circ}$.

\begin{figure*}[htb]
	\centering
	\includegraphics[width=0.9\textwidth]{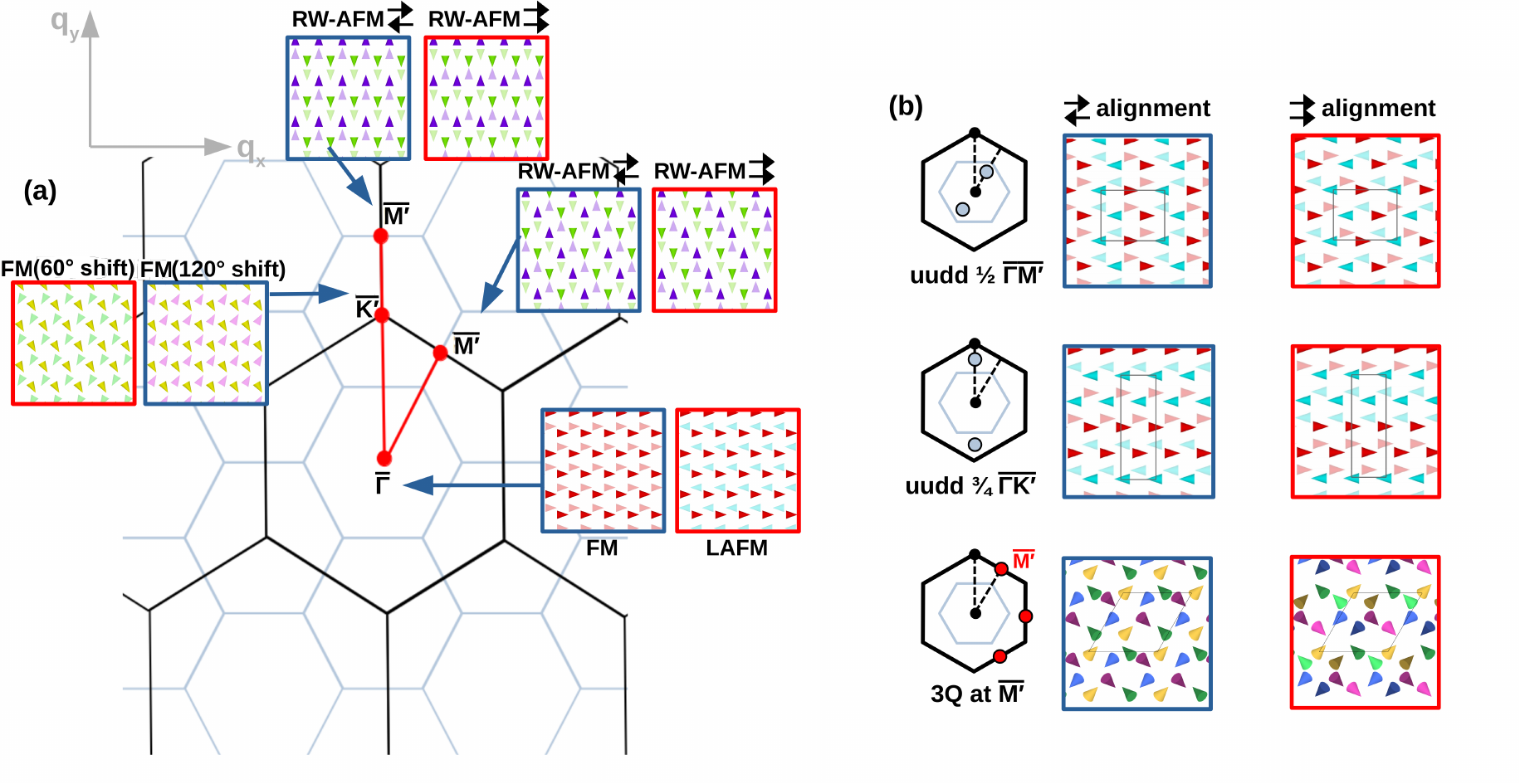}
	\caption{
		Sketch of the enlarged 2D hexagonal symmetry zone, a selection of spin spiral states arising at the high symmetry points and images of the prototypical multi-Q states of a magnetic bilayer. (a) depicts the enlarged 2D hexagonal symmetry zone in reciprocal space which needs to be taken into account for the calculation of spin spirals propagating simultaneously in both magnetic layers. For the high symmetry points spin structures are visualized for the case of $\rightleftarrows$ (blue) and  $\rightrightarrows$ (red) alignment of the layers at the 
		$\overline{\rm M}'$-point;
		bold (faded) colors indicate moments in the top (bottom) layer. The 2D BZ which suffices to treat spin spirals of a magnetic ML is depicted by a faded hexagon. (b) shows the definition and construction of the prototypical multi-Q states of a magnetic BL analogous to the one of a ML; for each coupling type of the layers, two collinear $uudd$ states and one non-collinear $3Q$ state are possible. For each spin structure the respective magnetic unit cell is indicated by a thin black line.} 
	\label{fig:BL_theory}
\end{figure*} 

For spin spirals propagating in two layers, the stacking sequence of the
magnetic layers is relevant.
As visible from Figs.~\ref{fig:BL_SS_setup}(c,d), in a magnetic bilayer the atomic positions of the
layers
are shifted against each other in the $xy$-plane,
due to the hexagonal stacking,
resulting in different distances between neighboring
interlayer
spins
compared to the ML (cf.~Fig.~\ref{fig:fittingfunctions}) \footnote{Note that the distance along $z$ does not matter in our setups since only spin spirals with $\mathbf{q}=(q_x, q_y,0)^\text{T}$ are being investigated in this work.}. With the introduction of the second layer, the rotational symmetry of the hexagonal ML around the $z$ axis in real space is partially broken, but from the atomic positions of both layers a new symmetric lattice (cf.~top views in Fig.~\ref{fig:BL_SS_setup}) with a smaller hexagonal cell can be constructed. 

In reciprocal space, the symmetry zone of spin spirals corresponding to such 
an effective honeycomb lattice is extended by a factor of $\sqrt{3}$ into every direction and rotated by an angle of 30$^{\circ}$ compared to the BZ of the hexagonal lattice of a magnetic ML (cf. Fig.~\ref{fig:BL_theory}(a)). 
Due to the extended 2D BZ
new high symmetry points occur which correspond to
spin structures of the bilayer and the alignment of spins between them. 
In addition, the computational effort in DFT increases significantly due to the longer high symmetry directions $\overline{\Gamma\text{M}'}$ and $\overline{\Gamma\text{K}'\text{M}'}$ containing a larger number of $\mathbf{q}$ values. 
The spin structures arising in spin spiral calculations
at the high symmetry points are determined by the alignment of the magnetic moments:
in the case of $\theta_{\rm T}=90^\circ$ and $\theta_{\rm B}=90^\circ$ (Fig.~\ref{fig:BL_SS_setup}(c))
the FM state emerges at the center of the symmetry zone
($\overline{\Gamma}$ point, see Fig.~\ref{fig:BL_theory}(a)), whereas it is the LAFM state 
for $\theta_{\rm T}=90^\circ$ and $\theta_{\rm B}=-90^\circ$ (Fig.~\ref{fig:BL_SS_setup}(d)).

\begin{figure}[htb]
	\centering
	\includegraphics[width=.9\linewidth]{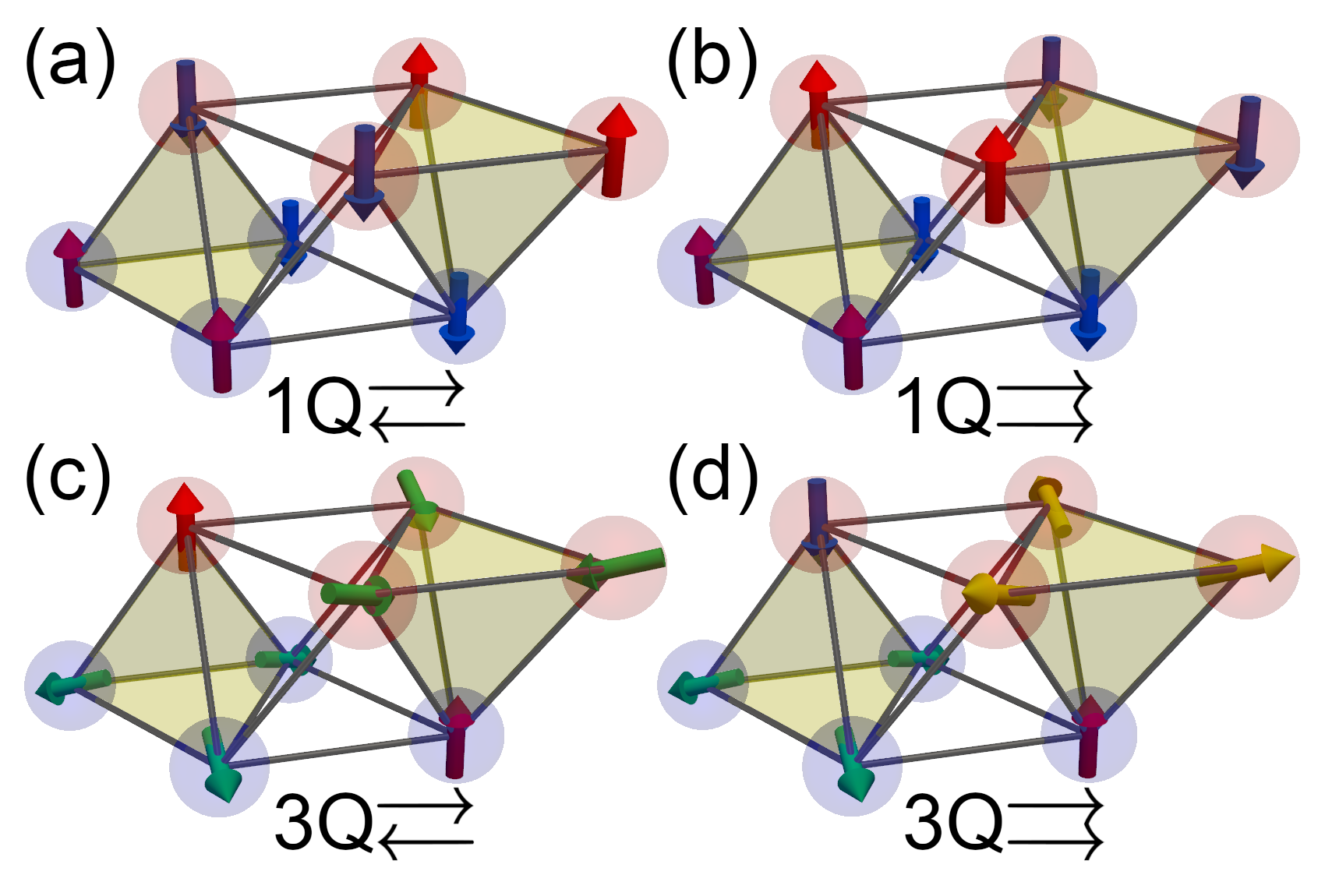}
	\caption{
		Illustration of the two types of (a,b) RW-AFM (1Q) and 
		(c,d) 3Q states in 
		a magnetic bilayer. 
		In every panel four magnetic atoms are shown
		in the top (red spheres) and in the bottom layer (blue spheres). Grey lines indicate the NN lattice connections within and between the top and bottom layer. 
		Arrows denote the magnetic moment directions.
		(a) RW-AFM$_{\rightleftarrows}$
		(1Q$_{\rightleftarrows}$) state: the layers exhibit a net antiparallel ($\rightleftarrows$) alignment, i.e.~one moment of a yellow tetrahedron lies in the hollow site of three moments of the other layer with one aligned parallel and two aligned antiparallel. 
		(b) RW-AFM$_{\rightrightarrows}$ (1Q$_{\rightrightarrows}$) state: magnetic moments of the upper layer are flipped with
		respect to the 1Q$_{\rightleftarrows}$ state
		and the layers are effectively aligned ferromagnetically 
		($\rightrightarrows$).
		(c) 3Q$_{\rightleftarrows}$ state: 
		magnetic moments form tetrahedron angles within and between the layers and the layers are net antiparallel aligned (see yellow tetrahedrons).
		(d) 3Q$_{\rightrightarrows}$ state: net parallel alignment of moments between the layers. In this state 
		tetrahedron angles occur only for NN moments
		within one layer and not
		between them.
	}
	\label{fig:BL_magnetisation}
\end{figure}

At the 
$\overline{\text{M}'}$-point at the zone boundary one obtains two types of 
RW-AFM states depending on the alignment between the layers.
In both magnetic configurations, there is a RW-AFM state within each of the 
two magnetic layers as for the ML case (cf.~Fig.~\ref{fig:ML_theory}). 
However, there are two possibilities of the net alignment of the moments
between the two layers:
in the net AFM alignment two of the three NN moments of bottom and top
layer are antiparallel 
(Fig.~\ref{fig:BL_magnetisation}(a))
while two of the three NN moments are parallel
for the net FM alignment  
(Fig.~\ref{fig:BL_magnetisation}(b)
and sketches in Fig.~\ref{fig:BL_theory}(a)). 
The RW-AFM state with a net AFM coupling between the layers
results at the $\overline{\text{M}'}$-point for spin spirals 
with 
$\theta_{\rm T}=90^\circ$ and $\theta_{\rm B}=90^\circ$
and the RW-AFM state with a net FM coupling between the layers
occurs for $\theta_{\rm T}=90^\circ$ and $\theta_{\rm B}=-90^\circ$.
To emphasize the net coupling between
the two layers, we denote these magnetic states of the bilayer
in the following
as the RW-AFM$_{\rightleftarrows}$ state 
and the RW-AFM$_{\rightrightarrows}$ state.

At the $\overline{\text{K}'}$-point corresponding to the center of the second periodic image of the hexagonal BZ of a magnetic ML one recovers
a FM state within each layer with moments rotated by 
$120^{\circ}$ or $60^{\circ}$
between the layers 
depending 
on the spin alignment between the layers.
Moreover, at the
$\frac{2}{3} \overline{\Gamma\text{M}'}$-point, corresponding to the $\overline{\text{K}}$-point of the BZ for a hexagonal monolayer, the N\'eel state is still present (not shown in Fig.~\ref{fig:BL_theory}(a)). 

The following discussions in our work focus on the two types of
RW-AFM 
states in the bilayer and the corresponding 3Q states
(Fig.~\ref{fig:BL_magnetisation}).
This motivates us to choose the following notation for spin spiral energy
dispersions based on the alignment of the moments between the
layers at the $\overline{\text{M}'}$ point of the BZ.
We denote the energy dispersions of the spin spirals as 
$E_{\rightleftarrows}(\mathbf{q})$ 
($\theta_{\rm T}=90^\circ, \theta_\text{B}=90^{\circ}$, Fig.~\ref{fig:BL_SS_setup}(c))
and $E_{\rightrightarrows}(\mathbf{q})$ 
($\theta_{\rm T}=90^\circ, \theta_\text{B}=-90^{\circ}$, Fig.~\ref{fig:BL_SS_setup}(d)). 
The
corresponding states at the $\overline{\text{M}'}$ point are
the RW-AFM$_{\rightleftarrows}$ state (Fig.~\ref{fig:BL_magnetisation}(a)) and the RW-AFM$_{\rightrightarrows}$ state
(Fig.~\ref{fig:BL_magnetisation}(b)), respectively, and
the energies are given by 
$E_{\text{RW-AFM}_{\rightleftarrows}}=E_{\rightleftarrows}(\mathbf{q}_{\overline{M}'})$ and
$E_{\text{RW-AFM}_{\rightrightarrows}}=E_{\rightrightarrows}(\mathbf{q}_{\overline{M}'})$ (Fig.~\ref{fig:BL_theory}(a)). 
At the $\overline{\Gamma}$-point, on the other hand, 
we obtain the FM and the 
layered AFM (LAFM) state (Fig.~\ref{fig:BL_theory}(a))
for the $\rightleftarrows$ and $\rightrightarrows$ alignment, respectively, and
the energies are given by $E_{{\rm FM}}=E_{\rightleftarrows}(0)$ and
$E_{{\rm LAFM}}=E_{\rightrightarrows}(0)$. 

For spin spirals propagating in both magnetic layers
energy contributions from intra- and interlayer pair-wise exchange arise which we denote as $E_{\parallel}(\mathbf{q})$ and
$E_{\perp}(\mathbf{q})$, respectively.
Based on the atomistic spin model, Eq.~(\ref{eq:H_BL}),
the energy contributions can be written as
\begin{equation}
	\begin{gathered}
		E_{\rightleftarrows}(\mathbf{q})=E_{\parallel}(\mathbf{q})+E_{\perp}(\mathbf{q})\\
		E_{\rightrightarrows}(\mathbf{q})=E_{\parallel}(\mathbf{q})-E_{\perp}(\mathbf{q}).\\
	\end{gathered}
	\label{eq:Energies_FM_AFM}
\end{equation}
Hence, by subtracting the DFT energy dispersion  $E_{\rightleftarrows}(\mathbf{q})$ from $E_{\rightrightarrows}(\mathbf{q})$ or adding the two 
dispersions
the curves of the interlayer exchange $E_{\perp}(\mathbf{q})$ or intralayer exchange 
$E_{\parallel}(\mathbf{q})$
can be filtered out. 

The enlargement of the magnetic phase space of spin spirals is accompanied by an increase of the number of prototypical multi-Q states. While there are two $uudd$ states and one 3Q state possible for a magnetic ML 
(cf.~Sec.~\ref{subsec:ML_dispersion}), twice the number can be constructed for the BL, based on the $\rightleftarrows$ and $\rightrightarrows$ alignment between the layers.
Indeed, these states are defined and constructed in analogy to those of the ML; 
$\mathbf{q}$=$\pm \frac{1}{2}\overline{\Gamma\text{M}'}$ and $\mathbf{q}$=$\pm \frac{3}{4}\overline{\Gamma\text{K}'}$ represent 90$^{\circ}$ spin spirals in each layer from which $uudd$ states can be formed by superposition. The first mentioned $\mathbf{q}$ value is identical to the one of the $uudd$-$\overline{\Gamma\text{K}}$ state of a magnetic ML, whereas the latter is located outside the first BZ (see Fig.~\ref{fig:BL_theory}(b)). The resulting $uudd$ states for both alignment types of the magnetic layers represent a combination of single-layer $uudd$ states with opposite
coupling between the layers (see sketches in Fig.~\ref{fig:BL_theory}(b)). 

\begin{figure*}
	\centering
	\includegraphics[width=0.85\textwidth]{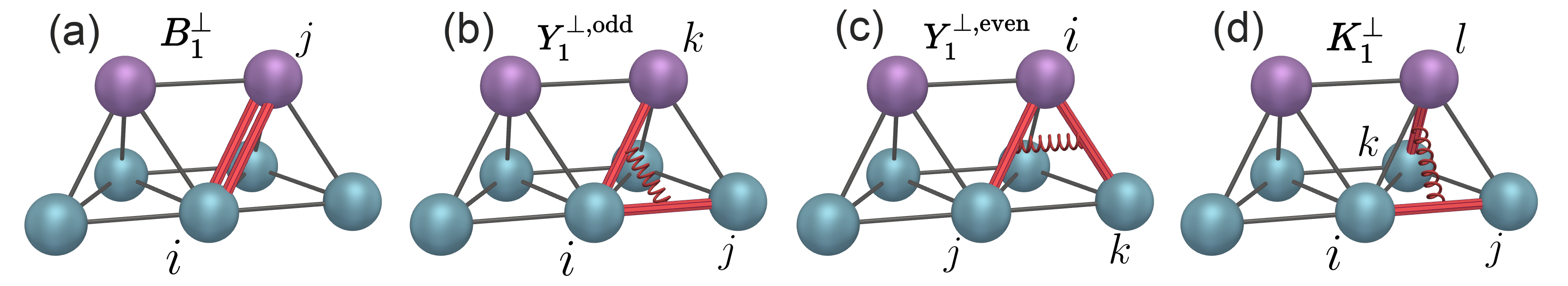}
	\caption{Illustration of nearest-neighbor interlayer higher-order exchange interactions in a
		hexagonal magnetic bilayer.
		In every panel only five atoms of the bottom layer (blue spheres)
		and two atoms of the top layer (purple spheres) are displayed
		with nearest-neighbor atoms connected by thin gray lines.
		The interacting atoms are marked with an index 
		($i, j, k, l$) matching to Eq.~(\ref{eq:H_HOI_perp}).
		(a) Biquadratic exchange interaction between the two layers with constant
		$B_{1}^{\perp}$.
		(b) Odd type of interlayer 4-spin 3-site interaction with
		three of the four spin moments in the two scalar
		products from the bottom layer and
		constant $Y_{1}^{\perp,{\rm odd}}$.
		(c) Even type of interlayer 4-spin 3-site interaction with evenly distributed spin moments in the scalar
		products from top and bottom layer and
		constant $Y_{1}^{\perp,{\rm even}}$.
		(d) Interlayer 4-spin 4-site interaction with spin
		moments on the corners of a tetrahedron and constant
		$K_{1}^{\perp}$.
		The 
		higher-order interactions are depicted through solid large red tubes, denoting the scalar products between spin moments, which are connected by red springs
		(cf.~Eq.~(\ref{eq:H_HOI_perp})). For clarity spin moments are not depicted in the sketches.     
		The three types of four-spin interactions are also illustrated
		in top view sketches in the Appendix (Fig.~\ref{fig:HOI_pairs_interlayer}) showing for
		one reference atom $i$ every pair of interacting moments.
	}
	\label{fig:BL_HOI}
\end{figure*}

Similarly, we can construct a 3Q state in the bilayer either from the 
RW-AFM$_{\rightleftarrows}$ states at the three equivalent $\overline{\text{M}'}$-points or from the
RW-AFM$_{\rightrightarrows}$ states (Fig.~\ref{fig:BL_theory}(b)). 
From the superposition
of the three RW-AFM$_{\rightleftarrows}$ states we obtain a 3Q state with a net
AFM coupling (Fig.~\ref{fig:BL_magnetisation}(c)), while a net FM coupling arises (Fig.~\ref{fig:BL_magnetisation}(d)) upon using the 
RW-AFM$_{\rightrightarrows}$ state.
We denote these two 3Q states as the 3Q$_{\rightleftarrows}$ and the
3Q$_{\rightrightarrows}$ state, respectively
(Fig.~\ref{fig:BL_magnetisation}(c,d)).
In both of these 3Q states, 
tetrahedron angles occur between adjacent spins within one of the layers.
However, only in the 3Q$_{\rightleftarrows}$ state (Fig.~\ref{fig:BL_magnetisation}(c))
tetrahedron angles 
are also spanned for nearest neighbor spins of different layers.
Therefore, we denote this spin structure as the 
ideal bilayer 3Q state.

Since the multi-Q states of the bilayer are obtained from a superposition of 
spin spiral (1Q) states as for the monolayer 
they are also energetically degenerate if we consider only
pair-wise Heisenberg exchange interactions,
$H_\text{BL}$ (Eq.~(\ref{eq:H_BL})).
In order to explain the total energy
differences found in DFT calculations as reported below, we need to take 
higher-order interactions into account. Considering HOIs in a magnetic bilayer introduces additional terms to the Hamiltonian. In the nearest neighbor approximation for a symmetric bilayer the following additional higher-order terms arise:

\begin{equation}
	\begin{split}
		H_{\rm HOI}^{\perp} = 
		&- \sum\limits_{\langle ij \rangle } B_{1}^{\perp} (\mathbf{m}_i^L \cdot \mathbf{m}_j^{L'})^2 \\
		&- \sum\limits_{\langle ijk \rangle} Y_{1}^{\perp, {\rm odd}} (\mathbf{m}_i^L \cdot \mathbf{m}_j^{L})(\mathbf{m}_i^L \cdot \mathbf{m}_k^{L'}) \\
		&- \sum\limits_{\langle ijk \rangle} Y_{1}^{\perp, {\rm even}}  (\mathbf{m}_i^L \cdot \mathbf{m}_j^{L'})(\mathbf{m}_i^L \cdot \mathbf{m}_k^{L'}) \\
		&- \sum\limits_{\langle ijkl \rangle } K_{1}^{\perp} (\mathbf{m}_i^L \cdot \mathbf{m}_j^{L})(\mathbf{m}_k^L \cdot \mathbf{m}_l^{L'}). \\
		\label{eq:H_HOI_perp}
	\end{split}
\end{equation}

Here the summations are performed over unit-cells 
$i, j, k$ and $l$ and the superscripts $L$ and $L'$ denote the top and bottom layer
($ L \in \{\text{B}, \text{T}\}$)
with $L \ne L'$
in all terms.

Four interlayer HOIs are introduced in Eq.~(\ref{eq:H_HOI_perp}) which are schematically depicted in Fig.~\ref{fig:BL_HOI}. 
The first term is the 
biquadratic exchange interaction 
between two nearest interlayer neighbors $i$ and $j$ of the top and bottom layer (Fig.~\ref{fig:BL_HOI}(a))
with the interaction constant $B_{1}^{\perp}$.
There are two different types of interlayer
4-spin 3-site interactions which occur since there are two 
possibilities to distribute three spin moments 
on the two scalar products. In the first type,
characterized by the interaction constant 
$Y_{1}^{\perp, {\rm odd}}$, there is a scalar
product of the two spin moments of the same layer $L$ 
and the second scalar product contains one spin
of the other layer $L'$
(Fig.~\ref{fig:BL_HOI}(b)). This term exhibits a change of sign
if the spin moments in one of the two layers are inverted, i.e.~for
the operation $\mathbf{m}_i^L$ to $-\mathbf{m}_i^L$.
Therefore, we denote it as an odd 4-spin 3-site 
interaction,
due to the odd distribution 3:1 on the layers $L:L'$.
The even interlayer 4-spin 3-site 
interaction denoted by the HOI constant 
$Y_{1}^{\perp, {\rm even}}$
is obtained if both scalar products
contain the spin moment of the other layer $L'$
(Fig.~\ref{fig:BL_HOI}(c)).
Note, that the summations are performed over triangles
of nearest neighbors within the two layers and
all permutations of the indices $i,j,k$.

Finally, we consider the interlayer 4-spin
4-site interaction in which the spin moments
are distributed on a tetrahedron (Fig.~\ref{fig:BL_HOI}(d)).
In this case all four sites are nearest neighbor atoms.
The interlayer 4-spin 4-site
interaction is characterized via the constant $K_{1}^{\perp}$
and is odd with respect to inverting the spin moments
either in the top or bottom layer.

For simplicity, we consider a symmetric bilayer in our
discussion of the interlayer HOI terms. For the Mn 
bilayer on the Ir(111) surface discussed below based
on DFT calculations, additional terms arise since the
two Mn layers are not equivalent. Thereby, we would 
obtain two additional 4-spin 3-site interaction constants
and one additional 4-spin 4-site interaction constant
depending on whether $L$ or $L'$ denote the top or
bottom layer in Eq.~(\ref{eq:H_HOI_perp}).    
While it is straightforward to generalize the spin model
in this way it is not required below
for the discussion of the intra- vs.~interlayer HOI
contributions to the DFT total energies.

The presented interlayer HOIs can be derived from a multi-band Hubbard model with spin $\geq1$ using fourth-order perturbations \cite{Hoffmann2020}. Our bilayer case would change the geometry of the model through new atomic sites featuring different hopping and onsite parameters ($t, U$). 
Introducing one atom of a second species into the 4-spin 3-site energy of a triangle of spins, Eq.~(36) of Ref.~\cite{Hoffmann2020}, would split the three terms up into the different interlayer 4-spin 3-site interactions ($Y^{\perp,\text{odd}}$, $Y^{\perp,\text{even}}$).

\section{Computational methods}
\label{sec:Comp_methods}
In order to investigate the magnetic exchange interactions of a freestanding Mn BL as well as Mn MLs and BLs on the Ir(111) surface, we resort to first-principles calculations based on DFT. Similar to previous work on ultrathin transition metal films on surfaces~\cite{Gutzeit2022, Gutzeit2023}, we make use of a combination of the full-potential linearized augmented plane wave (FLAPW) method implemented in the {\tt FLEUR} code~\cite{FLEUR,Kurz2004} and the projector augmented wave (PAW) method incorporated in the {\tt VASP} code~\cite{Bloechl1994, vasp2, Kresse1996, Kresse1999}. While all calculations on the freestanding Mn bilayer rely on the in-plane lattice constant of Ir determined within the local density approximation (LDA) of the exchange correlation functional taking a value of 2.70 {\AA}~\cite{Dupe2014}, the in-plane lattice constant of Ir obtained from the generalized gradient approximation (GGA) with a value of 2.75 {\AA}~\cite{vonBergmann2006} serves as a basis for Mn MLs and BLs on the Ir(111) substrate. 

In a first step, geometry optimizations for both fcc and hcp stacking of the Mn ML as well as the Mn BLs on Ir(111) have been performed via the {\tt VASP} code. Here, in accordance with experimental observations~\cite{Saxena2024}, only an hcp stacking of the subsurface Mn layer is considered, whereas for the top magnetic layer both stacking sequences are possible. Details on the structural relaxation as well as the resulting interlayer distances and energies of all relevant magnetic states can be found in Ref.~\cite{Saxena2024}. For the freestanding Mn bilayer no structural optimization has been carried out; instead, the equilibrium lattice parameter of the Ir bulk which amounts to 2.20 {\AA} is set for the interlayer distance between the two Mn layers.

Employing the geometry of the structural relaxation, we perform subsequent self-consistent calculations of the energy dispersion $E(\mathbf{q})$ of flat spin spirals for the freestanding Mn bilayer, the Mn ML and BL on Ir(111) using the {\tt FLEUR} code. For the Mn BL on the Ir substrate, we restrict ourselves to the calculation of the electronic and magnetic properties of just one stacking of the top Mn layer -- the hcp sequence -- since for this system the energetically lowest spin structure after geometry optimization -- the RW-AFM$_\rightleftarrows$
state (see also Ref.~\cite{Saxena2024}) -- turns out to be lower in energy than for an fcc stacking of the upper Mn layer. Due to symmetry, no significant change of the energy dispersion is expected 
upon choosing the fcc stacking sequence for the top Mn layer. These spin spiral calculations can be performed within the chemical unit cell by exploiting the generalized Bloch theorem~\cite{Kurz2004}. Exchange correlation effects are included in LDA by using the Vosko, Wilk and Nusair (VWN) functional~\cite{vwn}. While for the Mn ML on Ir(111) the computation of just one set of flat spin spirals with $\mathbf{q}$-vectors chosen along the two high symmetry directions of the 2D hexagonal BZ is sufficient for each stacking (see Sec.~\ref{subsec:ML_dispersion}), the situation turns out to be computationally much more demanding for the magnetic BL systems. As described in Sec.~\ref{subsec:BL_dispersion}, four sets of calculations are necessary for both the freestanding Mn bilayer and the hcp-Mn/hcp-Mn/Ir(111) film system in order to extract the relevant intra- and interlayer Heisenberg exchange constants. 

For these types of calculations, asymmetric films comprising nine Ir layers and one Mn ML (BL) on one side of the film are used. The muffin tin radii are chosen as 2.30 a.u.~for the 
Mn atoms, whereas a slightly larger value of 2.31 a.u.~is used for 
the non-magnetic 
Ir atoms. The number of $k$-points amounts to 1936 (2025) in the full 2D BZ and the cutoff for the basis functions is taken as $k_{\text{max}}$=4.1~a.u.$^{-1}$ (4.3~ a.u.$^{-1}$) for Mn/Ir(111) and hcp-Mn/hcp-Mn/Ir(111) (freestanding Mn bilayer). Starting from the self-consistent results for spin spirals, the energy contributions of the Dzyaloshinskii-Moriya interaction (DMI) for the film systems are calculated by including  SOC in first-order perturbation theory~\cite{Heide2009}.
The 
magnetocrystalline anisotropy energy
(MAE) is computed self-consistently \cite{Li1990} via the {\tt FLEUR} code as well by adding SOC to the RW-AFM state in the two-atomic unit cell. To obtain accurate results for this small quantity, the cutoff for the basis functions is increased to $k_{\text{max}}=4.3$~a.u.$^{-1}$ and the number of $k$-points to 8024 in the full 2D BZ.

For the Mn BL on Ir(111),
we obtain the total energies of all considered 
multi-Q superposition states
using the {\tt FLEUR} code. 
We used the same relaxed interlayer distances, exchange-correlation potential, cutoff for the basis functions and muffin tin radii as for the
spin spiral
energy dispersion to determine the energies of the $uudd$ states with respect to the $\rightleftarrows$ or $\rightrightarrows$
alignment
within their four-atomic unit cells per layer. %
Only the number of $k$-points needs to be adapted since the size of the BZ in reciprocal space changes:
for the multi-Q states, i.e. $uudd$ states along $\overline{\Gamma\text{M}}$ ($\overline{\Gamma\text{K$'$}}$) and $\overline{\Gamma\text{K}}$ ($\overline{\Gamma\text{M$'$}}$) direction 
and the 3Q state, 
we chose 288 $k$-points in the irreducible part of the 2D BZ.
The total energies for the
$uudd$ states
for the Mn ML on Ir(111) were also obtained
using the {\tt FLEUR} code with this setup.

For the self-consistent calculation of the total energies of the non-collinear 3Q state of the freestanding Mn bilayer and the 
Mn ML on Ir(111)
we resort to the {\tt VASP} code using the same relaxed interlayer distances as in the associated {\tt FLEUR} calculations, the LDA VWN potential~\cite{vwn} and the constrained magnetic moment approach. The energy cutoff is set to 300 eV in all cases and the 2D BZ is sampled by a 15$\times$15$\times$1 Monkhorst-Pack (MP) $k$-point mesh. The \textcolor{blue}{total} energies of the two $uudd$ states of the freestanding Mn bilayer are also obtained via {\tt VASP}; a 22$\times$28$\times$1 MP $k$-point mesh is applied for the multi-Q state along $\overline{\Gamma\text{M$'$}}$ direction and a 14$\times$44$\times$1 MP $k$-point mesh for the respective state along $\overline{\Gamma\text{K$'$}}$.

\section{Results}
\label{sec:Results}
\subsection{Mn monolayers on Ir(111)}
\begin{figure}[htb]
	\centering
\includegraphics[width=0.95\linewidth]{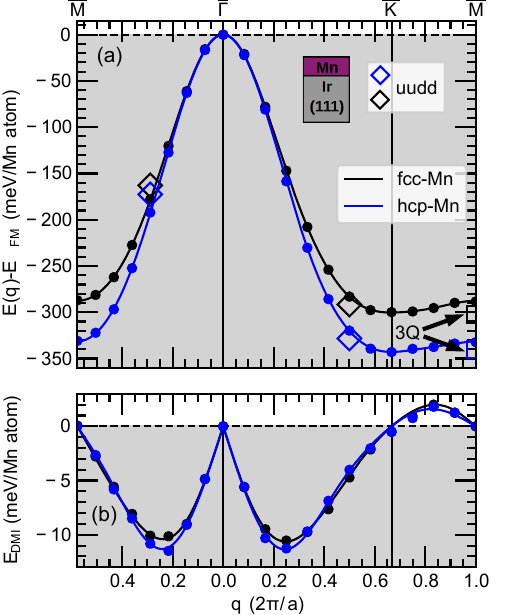}
	\caption{DFT calculated energy dispersion E($\mathbf{q}$) of flat cycloidal spin spirals for fcc- (black symbols) and hcp-stacked (blue symbols) Mn monolayers (MLs) on Ir(111). (a) Filled circles depict scalar-relativistic total DFT energies, while solid lines represent a fit to the model of intralayer pairwise Heisenberg exchange beyond nearest neighbors. The energies of the $uudd$ 
    states and the $3Q$ state are depicted as empty diamonds and squares at the absolute values of $\mathbf{q}$ of the constituting spin spiral states, respectively. (b) Total DMI contributions for cycloidal spin spirals of the two Mn stackings. The solid lines denote a fit to the Dzyaloshinskii-Moriya interaction. $E_{DM}<0$ ($>0$) indicates the preference of (counter)clockwise rotation of cycloidal spin spirals.}
	\label{fig:MnIr111_dispersion}
\end{figure}

Before turning to the more complex topic of magnetic bilayers, we first focus on the magnetic properties of Mn MLs on the Ir(111) surface in this section. Fig.~\ref{fig:MnIr111_dispersion}(a) displays the energy dispersion $E(\mathbf{q})$ of flat cycloidal spin spirals in the scalar-relativistic approximation, i.e.~neglecting the effect of SOC, for both fcc and hcp stacking of the Mn ML. While filled circles represent total energies calculated via DFT, the solid lines depict a fit to the model of pairwise Heisenberg exchange including eight shells of nearest neighbors. One immediately notices that 
$E(\mathbf{q})$ for
both stackings shows the same qualitative behavior with the RW-AFM state at the BZ boundary ($\overline{{\rm M}}$-point)
having a much lower energy (on the order of 290 to 330 meV/Mn atom) than the FM state at the $\overline{\Gamma}$-point. At the center of the BZ the magnetic moment of Mn in fcc stacking on Ir(111) (fcc-Mn) amounts to 3.35 $\mu_{\text{B}}$, the respective value of hcp-Mn is found to be given by 3.26 $\mu_{\text{B}}$.
In both cases the magnitude changes by only 2\% upon varying $\mathbf{q}$ along the high-symmetry directions of the 2D BZ.

As seen from Fig.~\ref{fig:MnIr111_dispersion}(a), for both Mn ML stackings on Ir(111) the N\'eel state at the $\overline{\rm K}$-point represents the state of lowest energy in the phase space of spin spirals. This can be understood by extracting the pairwise intralayer Heisenberg exchange constants from the fit of the energy dispersion (see Table~\ref{tab:MnIr111_J_NNHOI}): in both systems, $J_1^{\parallel}$ turns out to be the dominant exchange parameter mediating a strong antiferromagnetic exchange coupling between nearest neighbor (NN) spins. $J_2^{\parallel}$ is of negative sign as well with a ratio of 0.14$J_1^{\parallel}$ in case of fcc-Mn and 0.12$J_1^{\parallel}$ for hcp-Mn. As shown in Ref.~\cite{Kurz2000}, under the condition that either only $J_1^{\parallel}$ is present in an antiferromagnetic material or $J_2^{\parallel}$ prefers AFM coupling as well with a maximum strength of $\frac{1}{8}$$J_1^{\parallel}$, the N\'eel state emerges as the ground state due to geometric frustration of the exchange interactions on the triangular lattice. 
For fcc-Mn/Ir(111) the value of $J_2^{\parallel}$ exceeds the limit of 1/8=0.125, thus one needs to take into account the contribution of NNNN exchange $J_3^{\parallel}$. The condition then
becomes $J_2^{\parallel}>9/8 J_3^{\parallel} + 1/8 J_1^{\parallel}$ \cite{Kurz2001}, which is fulfilled by the exchange constants extracted for fcc-Mn.

The situation is different for the closely related system of a Mn ML on the Re(0001) surface~\cite{Spethmann2020}. Although in this case the Mn MLs are characterized by strong AFM intralayer exchange  between nearest neighbors as well, the energy minimum in the spin spiral phase space turns out to be the RW-AFM state due to exchange frustration of $J_1^{\parallel}$ and $J_2^{\parallel}$ having a large ratio of 0.23 up to 0.32 for the two Mn stackings.

\begin{table}[htb]
	\centering
	\caption{Heisenberg exchange constants for the first three nearest neighbors of fcc- and hcp-stacked Mn MLs on Ir(111), $J_1^{\parallel}$ to $J_3^{\parallel}$, as extracted from fitting the respective spin spiral DFT energy dispersion E($\mathbf{q}$), i.e. neglecting higher-order terms, and higher-order exchange constants $B_1^{\parallel}$, $K_1^{\parallel}$ and $Y_1^{\parallel}$ calculated from Eqs. (\ref{eq:HOI_energy1})-(\ref{eq:HOI_energy3}). All values are given in meV/Mn atom.}
	\label{tab:MnIr111_J_NNHOI}
	\begin{ruledtabular}
		\begin{tabular}{l c c c c c c }
			System& $J_1^{\parallel}$ &$J_2^{\parallel}$ &$J_3^{\parallel}$ & $B_1^{\parallel}$& $K_1^{\parallel}$ & $Y_1^{\parallel}$ \\
			\colrule
            fcc-Mn/Ir(111)&$-31.22$ &$-4.29$ &$-1.72$&$-3.19$ &$-1.22$ &$-3.02$ \\
            hcp-Mn/Ir(111)& $-36.78$ & ${-4.47}$ & ${-1.02}$ &${-3.19}$ &${-0.89}$ & $-3.50$ \\

			 \end{tabular} 
	\end{ruledtabular}
\end{table}	
Fig.~\ref{fig:MnIr111_dispersion}(b) further shows the total DMI contributions for every cycloidal spin spiral state calculated along the high symmetry directions of the 2D hexagonal BZ for the Mn/Ir(111) film system. Just like the scalar-relativistically computed DFT energies, they appear rather similar favoring a clockwise rotational sense of cycloidal spin spirals. Although the DMI parameter $D_1$ for nearest neighbors takes significant values of around 2.8 meV/Mn atom (see Table~\ref{tab:DMI_constants_MnIr111}), the effect of SOC becomes only relevant in the vicinity of the $\overline{\Gamma}$-point where the slope of the dispersion is steep. As expected, the DMI shows (nearly) no influence on the energetically close  N\'eel and RW-AFM states at the $\overline{\rm K}$- and $\overline{\rm M}$-point, respectively. 

Focusing on the last aspect of our analysis on Mn MLs on Ir(111), the prototypical multi-Q states, we notice that the total energies of the two collinear $uudd$ states and of the non-collinear 3Q state differ significantly from their corresponding single-Q (spin spiral) states (see Fig.~\ref{fig:MnIr111_dispersion}(a)). As a result, we obtain large values for the intralayer HOI  constants according to 
Eqs.~(\ref{eq:HOI_energy1}) to (\ref{eq:HOI_energy3}) (see Table~\ref{tab:MnIr111_J_NNHOI}); while the biquadratic term $B_1^{\parallel}$ and the 4-spin 3-site interaction strength $Y_1^{\parallel}$ are on the order of 3 meV/Mn atom, notably smaller values of about 1 meV/Mn atom result for the 4-spin 4-site term $K_1^{\parallel}$. 

Similar to the HOI constants calculated for Mn/Re(0001)~\cite{Spethmann2020}, all of them exhibit a negative sign. From Fig.~\ref{fig:MnIr111_dispersion}(a) it becomes obvious that these terms actually favor the 3Q superposition state over its constituting 1Q spin spiral state, the RW-AFM state, for both stackings of the Mn layer. 
    For fcc-Mn/Ir(111) the 3Q state gains 2.5~meV per Mn atom compared to the energetically lowest spin spiral state,
    i.e.~the N\'eel state, making it the lowest energy state.  
For hcp-Mn/Ir(111), in contrast, the 3Q state remains by 3 meV per Mn atom higher in energy than the N\'eel state consistent with the experimental observation
of the N\'eel state in this ultrathin film
system~\cite{Rodriguez2024}. 

\subsection{Mn bilayers on Ir(111)}
\begin{figure*}[htb]
	\centering
	\includegraphics[width=0.85\textwidth]{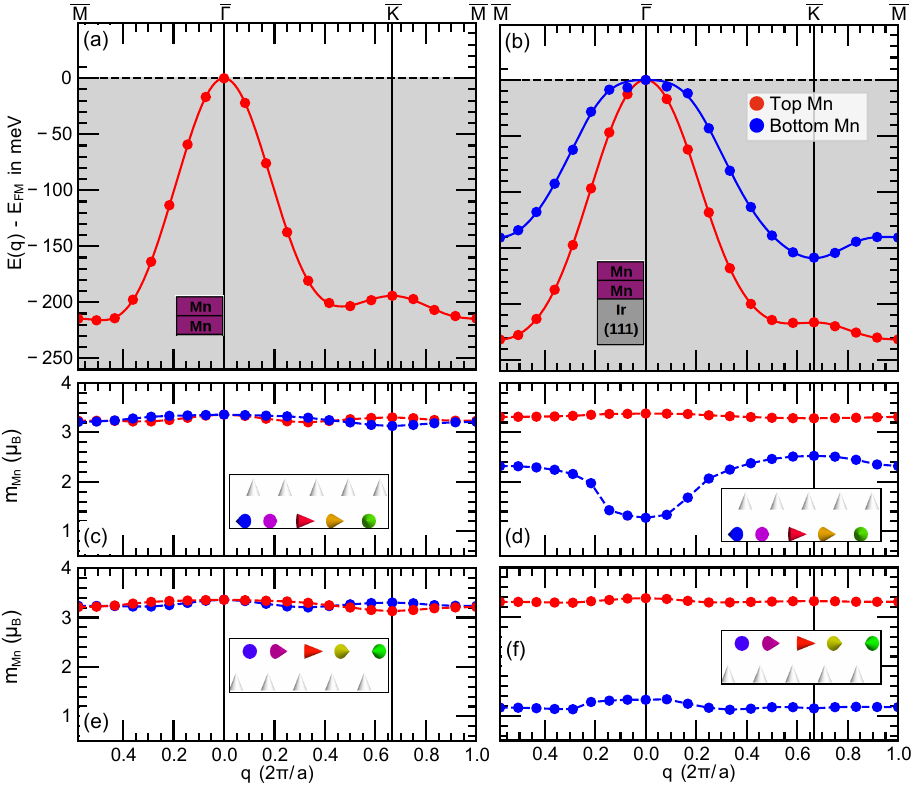}
	\caption{DFT calculated energy dispersion E($\mathbf{q}$) of a flat spin spiral propagating in (a) one of the equivalent magnetic layers
    of a freestanding Mn bilayer and (b) either in the top or in the
    bottom magnetic layer of a Mn BL on Ir(111). Total DFT energies in
    scalar-relativistic approximation, i.e.~neglecting SOC, are depicted by filled circles whereas solid lines denote fits to the model of intralayer pairwise Heisenberg exchange including up to eight nearest neighbor shells 
    (see Table~\ref{tab:Intralayer_exchange_MnBL}). (c-f) Magnetic moments of the Mn layers obtained for every $\mathbf{q}$ value calculated along the high symmetry lines of the 2D hexagonal BZ shown in (a) and (b). Note that (d) show Mn moments for a flat spin spiral propagating in the bottom (interface) magnetic layer of hcp-Mn/hcp-Mn/Ir(111) and (f) show Mn moments for a flat spin spiral propagating in the top (surface) magnetic layer of hcp-Mn/hcp-Mn/Ir(111) are displayed. For the free-standing Mn bilayer (c,e) the two layers are equivalent.
    Dashed lines between the DFT data points serve as a guide to the eye.}
	\label{fig:Intralayer_MnBL}
\end{figure*}

\textbf{Intralayer exchange.} Figs.~\ref{fig:Intralayer_MnBL}(a,b) show the energy dispersion E($\mathbf{q}$) neglecting SOC of flat spin spirals propagating only in one magnetic layer of a freestanding Mn bilayer and the Mn-BL/Ir(111) film system. Similar to Mn MLs on Ir(111), the two BL systems overall display a typical AFM coupling within the layers, with the RW-AFM state at the 
$\overline{\text{M}}$-point being energetically favored over the FM state at the center of the BZ by 130 to 230 meV/magnetic unit cell (u.c.), i.e.~per two Mn atoms. However, allowing the Mn bilayer to grow pseudomorphically on the Ir(111) surface has a drastic impact on the spin spiral state of lowest energy due to hybridization of Mn states with the non-magnetic substrate. 

For the freestanding Mn bilayer the global minimum of the energy dispersion is located at $q=|\mathbf{q}| \approx 0.51 \times 2 \pi /a$ along $\overline{\Gamma\text{M}}$ direction and hence only by 2 meV lower in energy than the RW-AFM state (Fig.~\ref{fig:Intralayer_MnBL}(a)). Note, that for this system the calculation of one spin spiral setup and one set of intralayer Heisenberg exchange constants (top or bottom magnetic layer) is sufficient since the two Mn layers are symmetry equivalent due to the vacuum on both sides. As revealed in Table~\ref{tab:Intralayer_exchange_MnBL}, the Mn bilayer is characterized by strong intralayer exchange frustration of AFM interactions between nearest and next-nearest neighboring spins. 

\begin{table*} [!htbp]
	\centering
    \caption{Intralayer Heisenberg exchange constants ($J_{i}^{\parallel}$) and Dzyaloshinskii-Moriya interaction (DMI) constants ($D_{i}^{\parallel}$) calculated via DFT for a freestanding Mn bilayer and the hcp-Mn/hcp-Mn/Ir(111) film system. The parameters have been determined individually for each Mn layer via a separate energy dispersion $E(\mathbf{q})$ as indicated in Figs.~\ref{fig:Intralayer_MnBL}(a,b) and Fig.~\ref{fig:DMI_MnBLIr111}(a) (Appendix~\ref{appendix:DMI}); the respective Mn layer is marked in bold letters. Note that in case of the freestanding Mn layer the calculation of the energy dispersion in one Mn layer suffices due to symmetry reasons and DMI contributions do not occur. Possible intralayer HOI terms are not taken into account here. $J_i^{\parallel}>0$ ($J_i^{\parallel}<0$) denotes ferromagnetic (antiferromagnetic) coupling and $D_i^{\parallel}>0$ ($D_i^{\parallel}<0$) the preference of a  clockwise (counterclockwise) rotation of 
    cycloidal spin spirals.
    All values are given in meV/magnetic unit cell (u.c.), i.e. per two Mn atoms.
		}
	\begin{ruledtabular}
		\begin{tabular}{lcccccccc}
			$\mathrm{System}$ & $J_{1}^{\parallel}$ & $J_{2}^{\parallel}$ & $J_{3}^{\parallel}$ & $J_{4}^{\parallel}$ & $J_{5}^{\parallel}$ & $J_{6}^{\parallel}$ & $J_{7}^{\parallel}$ & $J_{8}^{\parallel}$  \\
			\colrule
			Mn/Mn 
   & $-18.73$ & $-$8.17 & $-2.53$ & $-0.19$ & 0.34 & $-0.23$ & 0.07& $-0.04$ \\
			\textbf{hcp-Mn}/hcp-Mn/Ir(111) 
   & $-23.37$ & $-5.80$ & $-0.74$ & $-0.13$ &0.17 &$-0.13$& 0.15& 0.03 \\
           hcp-Mn/\textbf{hcp-Mn}/Ir(111)
           & $-18.91$ & 0.19 & 0.46 & 0.39 &0.52 &0.25&$-0.06$& 0.11\\
           \colrule
            & $D_{1}^{\parallel}$ & $D_{2}^{\parallel}$ & $D_{3}^{\parallel}$ & $D_{4}^{\parallel}$ & $D_{5}^{\parallel}$ \\ 
            \colrule
            \textbf{hcp-Mn}/hcp-Mn/Ir(111) & 0.33 & 0.22 & 0.11 & 0.07&$-0.06$ \\
            hcp-Mn/\textbf{hcp-Mn}/Ir(111) &1.82 &$-0.23$ &0.12 &$-0.10$ &$-0.26$ \\
		\end{tabular}
		\label{tab:Intralayer_exchange_MnBL}
	\end{ruledtabular}
\end{table*}

\begin{table*} [!htbp]
	\centering
    \caption{Interlayer Heisenberg exchange constants ($J_{i}^{\perp}$) calculated via DFT for a freestanding Mn bilayer and the hcp-Mn/hcp-Mn/Ir(111) film system. The parameters have been determined from the energy dispersions shown in Figs.~\ref{fig:Interlayer_MnBL}(c,d). $J_i^{\perp}>0$ ($J_i^{\perp}<0$) denotes ferromagnetic (antiferromagnetic) coupling between the moments of the two magnetic layers.
    All values are given in meV/magnetic u.c., i.e. per two Mn atoms.
		}
	\begin{ruledtabular}
		\begin{tabular}{lcccccccc}
			$\mathrm{System}$ & $J_{1}^{\perp}$ & $J_{2}^{\perp}$ & $J_{3}^{\perp}$ & $J_{4}^{\perp}$ & $J_{5}^{\perp}$ & $J_{6}^{\perp}$ & $J_{7}^{\perp}$ & $J_{8}^{\perp}$  \\
			\colrule
			  Mn/Mn 
   & 4.58 & $-6.20$ & $-4.39$ & 1.22 &1.16 &0.39& $-0.27$& $-0.34$ \\
           hcp-Mn/hcp-Mn/Ir(111)
           & $-19.65$ & $-3.57$ & $-2.93$ & $-0.27$ &0.41 &0.26&0.06&$-0.23$\\
		\end{tabular}
		\label{tab:Interlayer_exchange_MnBL}
	\end{ruledtabular}
\end{table*}

In the case of the Mn BL on Ir(111), spin spirals in the top 
magnetic layer 
(red symbols and curve in Fig.~\ref{fig:Intralayer_MnBL}(b))
exhibit a similar energy dispersion
as for the freestanding bilayer (cf.~Fig.~\ref{fig:Intralayer_MnBL}(a)). The dispersion clearly shows 
the RW-AFM state as the 1Q state of lowest energy. 
However, spin spirals propagating only in 
the bottom Mn layer (blue symbols and curve in 
Fig.~\ref{fig:Intralayer_MnBL}(b)), which is in direct contact with the Ir substrate, possesses a 120$^{\circ}$ N\'eel ground state similar to the case of the Mn ML on Ir(111) 
(cf.~Fig.~\ref{fig:MnIr111_dispersion}(a)). Again, this observation can be attributed to the ratio of the extracted intralayer
Heisenberg exchange constants (see Table~\ref{tab:Intralayer_exchange_MnBL}): the respective value between the AFM nearest and next-nearest neighbor constant $J_1^{\parallel}$ and $J_2^{\parallel}$, respectively, amounts to roughly $\frac{1}{4}$ for the top Mn layer thereby causing the formation of a RW-AFM ground state due to exchange frustration. The AFM coupling constant $J_1^{\parallel}$ turns out to be the dominant Heisenberg exchange parameter for the bottom Mn layer resulting in the emergence of the N\'eel state owing to geometric frustration on the hexagonal lattice,
$|J_2^{\parallel}|< 1/8|J_1^{\parallel}|$.

Fig.~\ref{fig:DMI_MnBLIr111}(a) in Appendix~\ref{appendix:DMI} further displays the energy contributions of the intralayer DMI computed for spin spiral states of the Mn/Mn/Ir(111) film system. As for the Mn ML on Ir(111), the preferred rotational sense of cycloidal spin spirals is clockwise for both Mn layers. As expected, the subsurface Mn layer experiences a larger impact of SOC compared to the top Mn layer which is well reflected in the DMI constant $D_1^{\parallel}$ for nearest neighbors (see Table~\ref{tab:Intralayer_exchange_MnBL}): with about 1.8 meV/u.c.~it is by a factor of about 6 larger than for the top layer.

Figs.~\ref{fig:Intralayer_MnBL}(c-f) display another interesting aspect of Mn bilayer systems: the variation of the magnetic moments of the individual layers with the spin spiral vector $\mathbf{q}$. While the values for the freestanding Mn BL are on the order of 3.30 $\mu_B$ and vary only little with $\mathbf{q}$ (Fig.~\ref{fig:Intralayer_MnBL}(c,e)), a different situation emerges for the Mn-BL/Ir(111) film system. Here, for the case of a flat spin spiral propagating only in the top Mn layer (Fig.~\ref{fig:Intralayer_MnBL}(f)), its moments are quite constant upon the variation of $\mathbf{q}$ taking a maximum deviation of 3\% from the value at the $\overline{\Gamma}$-point. On the other hand, the Mn moments of the bottom layer which are always oriented perpendicular to the rotation plane of the overlying spin spiral are strongly reduced to 1.33 $\mu_B$ with minimum values of 1.12 $\mu_B$ being taken for $\mathbf{q}$-vectors along $\overline{\Gamma\text{M}}$ (Fig.~\ref{fig:Intralayer_MnBL}(f)). 

If the flat spin spiral propagates only in the bottom Mn layer (Fig.~\ref{fig:Intralayer_MnBL}(d)), the moments in the top magnetic layer are again quite stable. Surprisingly, for the spin spiral in the bottom Mn layer 
an increase
of up to 98\% from the value of the $\overline{\Gamma}$-point occur: while the minimum of 1.27 $\mu_B$ is given exactly at the FM state, the N\'eel state at the $\overline{\text{K}}$-point holds the maximum of 2.52 $\mu_B$ (Fig.~\ref{fig:Intralayer_MnBL}(d)). This behaviour can be attributed to
the energy gain upon the formation of a magnetic moment for the ground state of the system.
Note, that variations in the magnitude of the magnetic moments are mapped effectively into the interaction constants of the classical Heisenberg model. Therefore, the spin model accounts for the fluctuations of the magnetic moment in the bottom Mn layer of the present system, resulting in a good fit of the energy dispersion in Fig.~\ref{fig:Intralayer_MnBL}(b).
Large variations of the magnetic moment for spin spirals along the high symmetry lines of the 2D BZ have also been reported for the bottom layer of a Mn BL on the W(110) surface~\cite{Schroeder2013}.
\begin{figure*}[htb]
	\centering
	\includegraphics[width=0.82\textwidth]{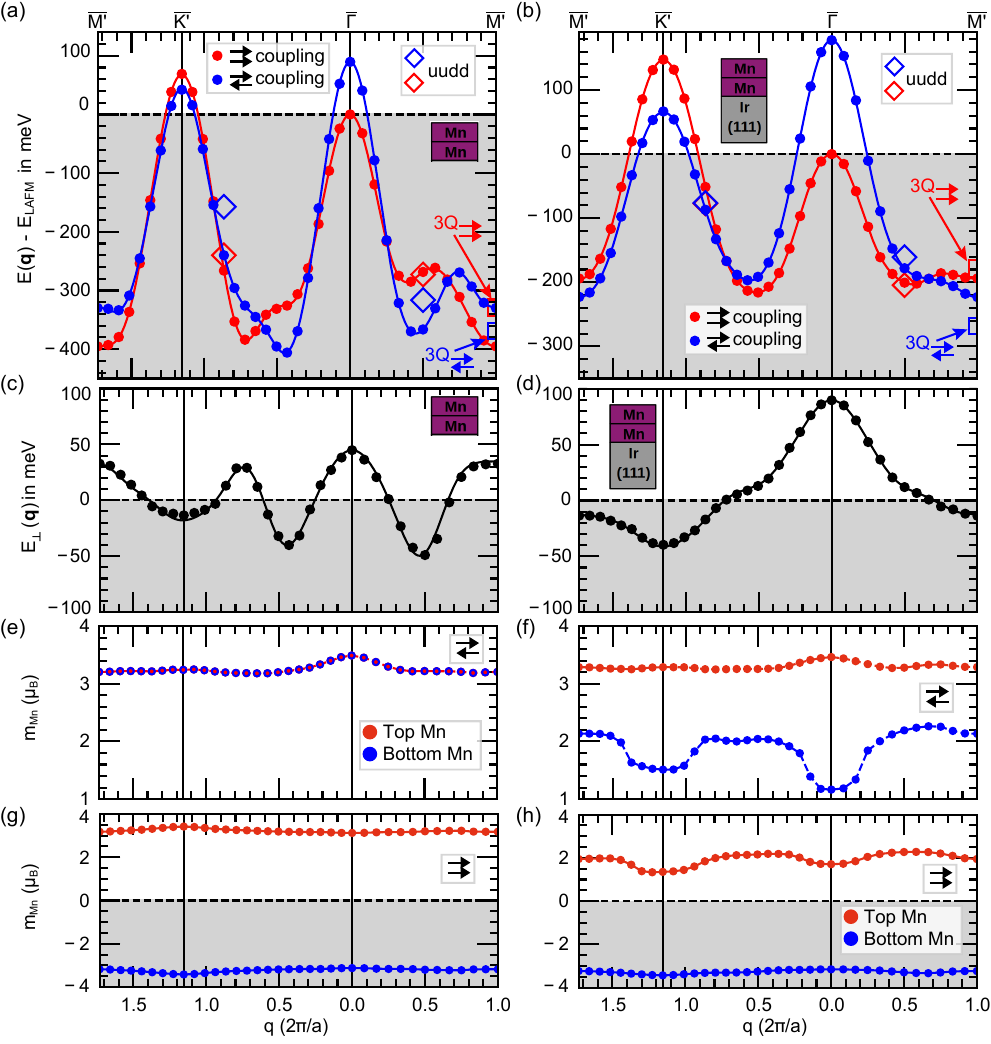}
	\caption{Energy dispersion E($\mathbf{q}$) 
    of flat spin spirals propagating in both magnetic layers of (a) a freestanding Mn bilayer and (b) the hcp-Mn/hcp-Mn/Ir(111) film system
    given with respect to the LAFM state. 
    Both $\rightleftarrows$ (blue) and $\rightrightarrows$ (red) alignment between the layers at the 
    $\overline{{\rm M}'}$-point 
    is taken into account. Total scalar-relativistic DFT energies are marked by filled circles of the respective color whereas solid lines denote fits to the Heisenberg model including intra- and interlayer exchange up to eight nearest neighbor shells. 
    Energies of $uudd$ 
    states and the 3Q states are depicted as empty diamonds and squares at the 
    $\mathbf{q}$ values of their constituting spin spiral states for 
    both alignments.
    Note, that the DMI contributions to spin spirals in the Mn BL on Ir(111)
    are given in the appendix (see Fig.~\ref{fig:DMI_MnBLIr111}(b)).
    (c,d) dispersion of the interlayer exchange calculated according to 
    $E_{\perp}(\mathbf{q})=\frac{1}{2}(E_{\rightleftarrows}(\mathbf{q})-E_{\rightrightarrows}(\mathbf{q}))$ 
    for the freestanding Mn bilayer and the hcp-Mn/hcp-Mn/Ir(111) film system, respectively. 
    (e-h) Magnetic moments of the Mn layers 
    for both spin alignments ($\rightleftarrows$ and $\rightrightarrows$)
    along the high symmetry lines of the enlarged hexagonal BZ. Note, that positive and negative values of the magnetic moments refer to the 
    alignment at the $\overline{\Gamma}$ point, i.e.~FM and LAFM state for $\rightleftarrows$ and $\rightrightarrows$ alignment, respectively.} 
	\label{fig:Interlayer_MnBL}
\end{figure*} 

\textbf{Interlayer exchange.}
The energy dispersions of flat spin spirals propagating only in one of
the magnetic layers as presented in the previous section are perfectly suited
to obtain the intralayer exchange constants and thereby provide 
valuable insight into the magnetism of the bilayer. 
However, the interlayer exchange contributions are missing and 
these DFT results cannot predict the possible ground state of our Mn bilayer on Ir(111). Therefore, additional calculations for spin spirals propagating simultaneously in both layers are required. Figs.~\ref{fig:Interlayer_MnBL}(a) and (b) illustrate a comparison of the DFT data of the just mentioned spin spiral setups for the freestanding Mn bilayer and the Mn BL on Ir(111).
Total DFT energies for a $\rightleftarrows$ ($\rightrightarrows$) alignment are denoted by blue (red) circles and solid lines of the respective color indicate fits to the model of pair-wise Heisenberg exchange including both intra- and interlayer exchange. Note that these lines just serve as a guide to the eye and that the exchange parameters of both systems listed in Table~\ref{tab:Intralayer_exchange_MnBL} and~\ref{tab:Interlayer_exchange_MnBL} are determined separately from Figs.~\ref{fig:Intralayer_MnBL}(a,b) and~\ref{fig:Interlayer_MnBL}(c,d) \footnote{In principle, both intra- and interlayer pairwise Heisenberg exchange constants could be determined simultaneously from a fit to the energy dispersion of $\rightleftarrows$ or $\rightrightarrows$ coupled magnetic layers. However, this only works for a symmetric bilayer with the same atom type in both layers and without substrate since the fitting routine cannot clearly assign the intralayer exchange to the corresponding layer.}. The energy dispersion of the $\rightleftarrows$ aligned layers, $E_{\rightleftarrows}(\mathbf{q})$,
is shifted in both plots with respect to the energy at the $\overline{\Gamma}$-point in the $\rightrightarrows$ alignment, $E_{\rightrightarrows}(\mathbf{q}=0)$, by the difference
$\Delta E$=$E_{\text{LAFM}}-E_{\text{FM}}$ obtained from the respective collinear calculations.

From Figs.~\ref{fig:Interlayer_MnBL}(a,b) one clearly sees that the general trend of the energy dispersion of the freestanding Mn bilayer is preserved upon placing it on the Ir(111) surface. $\Delta E$=$E_{\text{LAFM}}-E_{\text{FM}}$ takes a large negative value of about $-90$~meV/magnetic u.c.~in case of the freestanding
Mn bilayer and about
$-180$~meV/magnetic u.c.~for the ultrathin film system,
showing
a strong AFM tendency in both systems consistent with previous observations on the intralayer spin spiral dispersion. The FM state with the same orientation of the magnetic moments between the two layers (see Fig.~\ref{fig:BL_theory}(a)) arising at the $\overline{\Gamma}$-point 
($E_{{\rm FM}}=E_{\rightleftarrows}(0)$) represents the global energy maximum of the enlarged hexagonal symmetry zone in
both cases.
Another local energy maximum can be found at the high symmetry $\overline{\text{K}'}$-point which corresponds to the $\overline{\Gamma}$-point of the second periodic image of the original hexagonal BZ (see Fig.~\ref{fig:BL_theory}(a)), 
corresponding to a FM state within each layer with moments rotated by %
$120^{\circ}$ or $60^{\circ}$
between the layers 
depending
on the spin alignment between
the two layers.

However, the global and local energy minima of the spin spiral dispersion in the enlarged hexagonal symmetry zone turn out to be different for the two systems mainly due to hybridization between the magnetic bilayer and the non-magnetic Ir substrate. As visible from Fig.~\ref{fig:Interlayer_MnBL}(a), the state of lowest energy for the freestanding Mn bilayer is a 135$^{\circ}$ spin spiral in the $\rightleftarrows$ aligned layers at $|\mathbf{q}| \approx 0.43 \times 2 \pi /a$ along $\overline{\Gamma\text{K}'}$ (blue curve)
and another local minimum is given by the 
RW-AFM$_{\rightrightarrows}$ 
state at the boundary of the enlarged symmetry zone ($\overline{\text{M}'}$-point, red curve). On the other hand, for the Mn BL on Ir(111) the inclusion of DMI (see Fig.~\ref{fig:DMI_MnBLIr111}(b) in the Appendix)
does not alter the %
computed energetically lowest 1Q state which is the RW-AFM$_{\rightleftarrows}$ 
($\overline{\text{M}'}$-point, blue curve) characterized by a net AFM coupling between the magnetic moments of the two layers (see Fig.~\ref{fig:Interlayer_MnBL}(b)) \footnote{Note that due to inversion symmetry no DMI contributions for the freestanding Mn bilayer are possible}.

Figs.~\ref{fig:Interlayer_MnBL}(c) and (d) show the dispersion of the interlayer Heisenberg exchange calculated according to
$E_{\perp}(\mathbf{q})=\frac{1}{2}(E_{\rightleftarrows}(\mathbf{q})-E_{\rightrightarrows}(\mathbf{q}))$
for the two systems under discussion. Here, the subtle differences from the energy dispersion of $\rightleftarrows$ and $\rightrightarrows$ aligned layers between the simple freestanding Mn bilayer and the more complex film system as mentioned above become more apparent.
The sign of $E_{\perp}(\mathbf q)$ gives the tendency of aligning the layers to each other, i.e.~in the
$\rightrightarrows$ ($\rightleftarrows$)
alignment for a positive (negative) contribution of $E_{\perp}(\mathbf q)$.
For the freestanding Mn bilayer (Fig.~\ref{fig:Interlayer_MnBL}(c)),
two deep energy minima arise along the two high symmetry directions of the enlarged hexagonal symmetry zone with another shallow local
minimum showing up in the vicinity of the $\overline{\text{K}'}$-point.
The interlayer exchange energy of the Mn BL on Ir(111) 
(Fig.~\ref{fig:Interlayer_MnBL}(d))
is varying less, no local minima are observed, and the global minimum forms at the $\overline{\text{K}'}$-point.

This different behavior can 
be explained with the interlayer Heisenberg exchange constants obtained from a fit to the energy dispersions
(see Table~\ref{tab:Interlayer_exchange_MnBL}).
Surprisingly, for the freestanding Mn bilayer $J_1^{\perp}$ mediates a FM coupling between neighboring spins of the two  magnetic layers competing with AFM interactions of second and third nearest neighbors, $J_2^{\perp}$ and $J_3^{\perp}$, respectively, which are also on the same order of magnitude thereby revealing a strong interlayer exchange frustration. For the Mn BL on the Ir(111)  surface, however, the AFM interlayer exchange constant $J_1^{\perp}$ with a value of $\approx -20$~meV/magnetic u.c.~is the dominant exchange parameter resulting in strong AFM bonds between the magnetic layers. Therefore, the LAFM state is favored over the FM state at the
$\overline{\Gamma}$ point, which leads to $E_\perp(0)>0$ and at the $\overline{{\rm M}}'$ point the 
RW-AFM$_\rightleftarrows$ state is preferred and thus $E_\perp(\mathbf{q}_{\overline{{\rm M}}'})<0$.

Turning our focus to the magnetic moments of spin spiral states in two magnetic layers of both $\rightleftarrows$ and $\rightrightarrows$ alignment (Figs.~\ref{fig:Interlayer_MnBL}(e-h)), we observe a similar behavior as for the case of the intralayer exchange.
Upon variation of $\mathbf{q}$ along the high-symmetry directions
of the BZ, the spin moment remains quite stable on the order of 
about 3 to 3.2 $\mu_B$
with increasing values up to 3.4 $\mu_B$ near the $\overline{\Gamma}$-
and the $\overline{{\rm K}}'$ point
for %
spin spirals with $\rightleftarrows$ and $\rightrightarrows$
in the freestanding Mn bilayer, respectively (Figs.~\ref{fig:Interlayer_MnBL}(e,g)).
A similar trend is observed in 
the top Mn layer of the Mn BL on Ir(111) for which the maximum fluctuation with $\mathbf{q}$ amounts to 9\% for both alignments
(Figs.~\ref{fig:Interlayer_MnBL}(f,h)).
However, as in the case of a flat spin spiral propagating only in one magnetic layers (cf.~Figs.~\ref{fig:Intralayer_MnBL}(b,d)), a different situation emerges for the bottom Mn layer which is
in direct contact with the Ir(111) surface.
For the $\rightleftarrows$ alignment of the Mn layers
(Fig.~\ref{fig:Interlayer_MnBL}(f)),
the moment at the $\overline{\Gamma}$-point,
i.e.~in the FM state,
is not only strongly reduced to 1.16 $\mu_B$, but also varies by 94\% for $\mathbf{q}$-vectors along the high symmetry lines of the enlarged symmetry zone for spin spirals with the maximum value of 2.26 $\mu_B$ being taken at $|\mathbf{q}| = 0.66 \times 2 \pi /a$ along $\overline{\Gamma \text{M}'}$ (corresponding to a N\'eel state in each magnetic layer). A similar observation is made for the bottom Mn layer in 
the $\rightrightarrows$ alignment between the two magnetic layers
(Fig.~\ref{fig:Interlayer_MnBL}(h)).
Note, that the  $\overline{\Gamma}$ point corresponds to the LAFM state in this alignment
based on our definition. 

\textbf{Multi-Q states.} 
Increasing the explored part of the magnetic phase space of the freestanding Mn bilayer and the Mn BL on Ir(111) to the prototypical multi-Q states, we obtain further surprising results. As recognizable in Fig.~\ref{fig:Interlayer_MnBL}(a), the $uudd$ states along the $\overline{\Gamma {\rm K}}'$ and the $\overline{\Gamma {\rm M}}'$ direction
for the $\rightrightarrows$ alignment of the 
freestanding Mn bilayer
are quite close in energy to 
the corresponding 90$^{\circ}$ spin spirals. However, the energies of the respective superposition states for the $\rightleftarrows$ alignment deviate significantly from those of the corresponding 1Q states (see 
Table~\ref{tab:MultiQ_Ediff_bilayer} for exact values). 

For the more complex film system of a Mn bilayer on Ir(111)
all four $uudd$ states turn out to be energetically close to their constituents (see 
Fig.~\ref{fig:Interlayer_MnBL}(b) and Table~\ref{tab:MultiQ_Ediff_bilayer}).
In contrast to the $uudd$ states and 90° spin spirals, the multi-Q states at the
$\overline{{\rm M}'}$ point show a 
distinctively different trend.
For a net FM alignment of spins in the two layers,
the 3Q$_{\rightrightarrows}$ state
(cf.~Fig.~\ref{fig:BL_magnetisation}(d))
rises in energy with respect to its respective 1Q state, the RW-AFM$_{\rightrightarrows}$ state.
However, for a net AFM alignment, i.e.~for the 3Q$_{\rightleftarrows}$ state (cf.~Fig.~\ref{fig:BL_magnetisation}(c)),
it is energetically favored over the RW-AFM$_{\rightleftarrows}$ state. This effect is particularly pronounced in Mn/Mn/Ir(111)
(Fig.~\ref{fig:Interlayer_MnBL}(b))
with the 3Q$_{\rightleftarrows}$ state gaining
45 meV/magnetic u.c.~compared to its corresponding 1Q state,
i.e.~the RW-AFM$_{\rightleftarrows}$ state. 

\textbf{Higher-order exchange interactions.}
One can understand this surprising behavior by taking
into account
the interlayer HOI terms introduced in Eq.~(\ref{eq:H_HOI_perp}) of section~\ref{subsec:BL_dispersion}.
The Hamiltonian is of the form:
\begin{equation}
    H = H_{\text{s-BL}} + H_{\text{HOI}}^{\parallel} + H_{\text{HOI}}^{\perp}
\end{equation}
with the 
separate terms  
corresponding to the pair-wise Heisenberg exchange of a symmetrical bilayer, $H_{\text{s-BL}}$,
the intralayer HOIs within each layer,
$H_{\text{HOI}}^{\parallel}$,
and the interlayer HOIs of the bilayer,
$H_{\text{HOI}}^{\perp}$
given in Eqs.~(\ref{eq:H_BL_sym}),
(\ref{eq:H_HOI_par}), and
(\ref{eq:H_HOI_perp}), respectively. We limit our further discussions
to a Hamiltonian of a symmetric bilayer for easier comprehensibility, however,
generalizing the HOIs to an asymmetrical bilayer is straightforward.
For the symmetric bilayer, the intralayer %
pair-wise
and higher-order exchange constants of the top and bottom layer coincide and the interlayer higher-order interactions are symmetric. This reduces significantly the number of terms one needs to distinguish.

The energies of the RW-AFM$_{\rightleftarrows}$ and
of the RW-AFM$_{\rightrightarrows}$ state,
 $E_{\text{RW-AFM}}^{\rightleftarrows}$ 
 and $E_{\text{RW-AFM}}^{\rightrightarrows}$, 
 are given by
\begin{equation}
\begin{split}
E_{\text{RW-AFM}}^{\rightleftarrows,\rightrightarrows} 
= & 2 J_1^{\parallel} \pm  J^{\perp}_1
                    - 6 B_1^{\parallel}  
                    + 4 Y_1^{\parallel}
                    - 12 K_1^{\parallel}
                    \\
                    &
                    - 3 B^{\perp}_1                  
                    \pm 2 Y_{1}^{\perp,{\rm odd}}
                    + 2 Y_{1}^{\perp,{\rm even}}
                    \mp 6 K_1^{\perp}
 \label{eq:RW-AFM_energy_contributions}
 \end{split}
\end{equation}
showing the energy contributions from the discussed NN exchange interactions.
Note, that the change of spin alignment between the two layers introduces a sign flip of oddly distributed interactions.

The energy contributions to the respective 3Q$_{\rightleftarrows}$ and the 3Q$_{\rightrightarrows}$ state are given by
\begin{equation}
\begin{split}
E_{\text{3Q}}^{\rightleftarrows,\rightrightarrows}  = & 2 J_1^{\parallel}
                     \pm J^{\perp}_1
                    - \frac 2 3 B_1^{\parallel}
                    - \frac 4 3 Y_1^{\parallel}
                    - \frac 4 3 K_1^{\parallel}
                    \\
                    &
                     - \frac 13 B^{\perp}_1 
                    \mp \frac 2 3 Y_{1}^{\perp,{\rm odd}}   
                    - \frac 2 3 Y_{1}^{\perp, {\rm even}}
                    \mp \frac 2 3 K_1^{\perp}\text{.}
 \label{eq:3Q_energy_contributions}
 \end{split}
\end{equation}
The first point to notice upon comparing Eqs.~(\ref{eq:RW-AFM_energy_contributions}) and (\ref{eq:3Q_energy_contributions})
is that the bilinear exchange interaction constants $J_1^{\parallel}$ and $J^{\perp}_1$ show the energetic degeneracy of the RW-AFM and corresponding 3Q states in the pair-wise Heisenberg exchange Hamiltonian $H_{\text{s-BL}}$. 
If we consider the energy difference between these states the pair-wise exchange
terms cancel and we obtain
\begin{equation}
\begin{split}
&\Delta E_{\text{3Q}}^{\rightleftarrows,\rightrightarrows}=
E_{\text{3Q}}^{\rightleftarrows,\rightrightarrows} - E_{\text{RW-AFM}}^{\rightleftarrows,\rightrightarrows} = \\
&\frac{8}{3}   \left(
                    2 B_1^{\parallel}
                    - 2 Y_1^{\parallel}
                    + 4 K_1^{\parallel}+  B^{\perp}_1 
                    \mp  Y_{1}^{\perp,{\rm odd}}   
                    - Y_{1}^{\perp, {\rm even}}
                    \pm 2 K_1^{\perp}
            \right).\\
 \label{eq:3Q-RW-AFM_energy_contributions}
 \end{split}
\end{equation}

\begin{figure*}
\centering
\includegraphics[width=\textwidth]{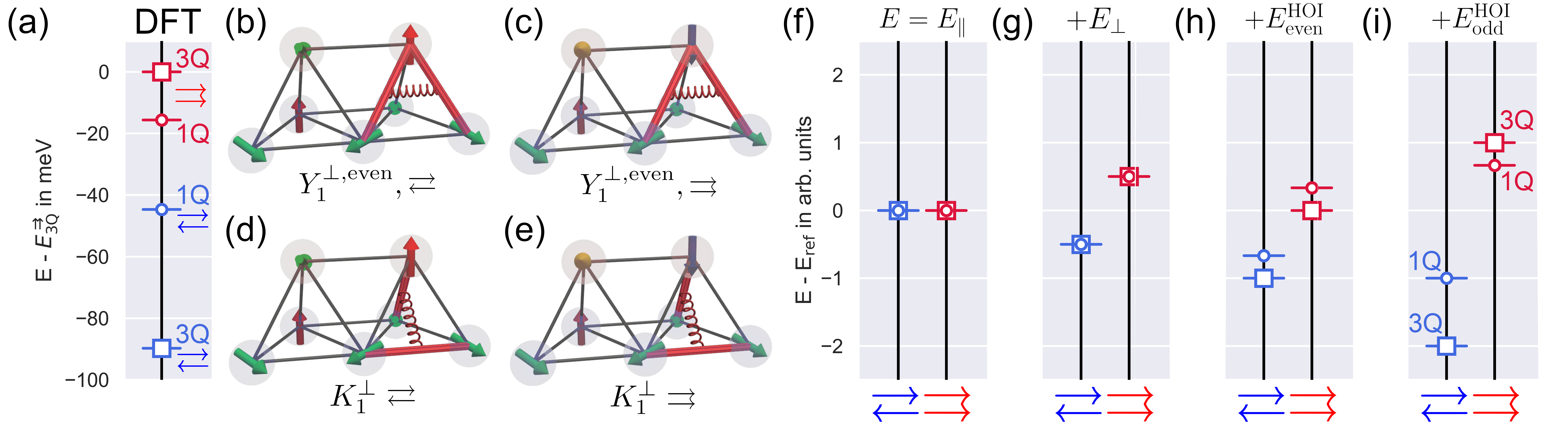}
\caption{
(a) DFT total energy differences for the Mn bilayer on Ir(111) at the $\overline{\text{M}'}$-point of the Brillouin zone with respect to the 3Q$_\rightrightarrows$
    state in meV per 2 Mn atoms. The energies of the RW-AFM (1Q, circle) and the 3Q state (square) are shown for 
    $\rightleftarrows$ (blue) and $\rightrightarrows$ (red) alignment of the spins in the two layers (for states cf.~Fig.~\ref{fig:BL_magnetisation}). 
    (b,c) The evenly distributed interlayer 4-spin 3-site interaction $Y_1^{\perp, {\rm even}}$ is displayed schematically for the 3Q state in $\rightleftarrows$
    and $\rightrightarrows$ alignment, respectively.
    The same spin structures are shown in (d,e) with the oddly distributed $K_1^{\perp}$ interaction. %
    (f-i) Energies of 1Q (circle) and 3Q (square) states in $\rightleftarrows$ (blue) and $\rightrightarrows$ (red) alignments are shown qualitatively on an arbitrary energy scale. From (f) to (i) the energy contributions of 
    Eq.~(\ref{eq:eff_energy_contributions}) are shown.  (f) Only bilinear intralayer exchange contributes with $E_{\parallel} = -1/2$.
    (g) Bilinear interlayer exchange is added with a value of $E_{\perp} = -1/2$.
    (h) Even intra- and interlayer HOI are added with a value of $E_{\mathrm{even}}^{\mathrm{HOI}} = -1/2$
    for the 3Q states.
    (i) Odd interlayer HOI is added setting $E_{\mathrm{odd}}^{\mathrm{HOI}} = -1$
    for the 3Q states.
    The HOI energy contributions ($E_{\mathrm{even}}^{\mathrm{HOI}}, E_{\mathrm{odd}}^{\mathrm{HOI}}$) to the 1Q states in (h,i) are %
    arbitrarily chosen to be $1/3$ of the 
    chosen values for the 3Q states.
    }
    \label{fig:energydiff_HOIs}
\end{figure*}

HOI lift the degeneracy of 1Q and 3Q states, as seen from the difference between $E_{\text{3Q}}^{\rightleftarrows,\rightrightarrows}$ and $E_{\text{RW-AFM}}^{\rightleftarrows,\rightrightarrows}$ 
and only contributions from intra- and
interlayer HOI remain in Eq.~(\ref{eq:3Q-RW-AFM_energy_contributions}).
Note, that the pair-wise exchange terms also cancel if we go beyond the NN approximation and consider inter- and intralayer
exchange up to an arbitrary neighbor shell.
We have performed similar calculations for the
90° spin spirals along the two high symmetry
directions
and the corresponding $uudd$ states and the
results
are given in Eqs.~(\ref{eq:1QGM_energy_contributions})
to (\ref{eq:UDGK_energy_contributions}) in
Appendix \ref{app:E_uudd_90deg}.

The DFT total energies
of the RW-AFM (1Q) and 3Q states
are displayed for the Mn bilayer on Ir(111) for
both $\rightrightarrows$ and $\rightleftarrows$
spin alignment
in Fig.~\ref{fig:energydiff_HOIs}(a) with respect to the
energy of the 3Q$_{\rightrightarrows}$ state. 
Note, that these energies can also be seen
at the $\overline{\rm M}'$ point
in Fig.~\ref{fig:Interlayer_MnBL}(b).
The RW-AFM$_{\rightleftarrows}$ (1Q$_{\rightleftarrows}$) state with a net antiferromagnetic coupling
between the layers is clearly favorable by about 25~meV/unit cell than the 
RW-AFM$_{\rightrightarrows}$ (1Q$_{\rightrightarrows}$) state with ferromagnetic coupling. However,
the energy gain of the 3Q state with respect to its corresponding 1Q state 
is very different for the two spin alignments between the layers.
While the 3Q$_\rightrightarrows$ state for the case of net ferromagnetic alignment lies roughly 15 meV/uc above the RW-AFM$_\rightrightarrows$, the 3Q$_{\rightleftarrows}$ state shows a large energy gain
of 45 meV/uc
with respect to the corresponding RW-AFM$_{\rightleftarrows}$ state in the net antiferromagnetic alignment (for the four spin structures see Fig.~\ref{fig:BL_magnetisation}).
    
This unexpected result can be understood based on Eq.~(\ref{eq:3Q-RW-AFM_energy_contributions}). All intralayer HOI terms
($B_1^\parallel$, $Y_1^\parallel$, $K_1^\parallel$) and two of the interlayer HOI terms
($B_1^\perp$, $Y_1^{\perp, {\rm even}}$) contribute in the same way to both energy differences.
However, two of the interlayer HOI terms ($Y_1^{\perp,{\rm odd}}$, $K_1^{\perp}$) have
opposite signs for the two 
spin alignments. The different behavior of these HOI terms is explained
from their symmetry (even or odd) if the spin moments of one of the two layers are flipped,
i.e.~switching between the $\rightleftarrows$ and the $\rightrightarrows$ state
(going from Fig.~\ref{fig:energydiff_HOIs}(b) to (c) or from Fig.~\ref{fig:energydiff_HOIs}(d) to (e)).
All intralayer HOI terms are even, i.e.~there is no change of sign in the energy terms,
as well as the biquadratic interlayer HOI and the even 4-spin 3-site interaction
(Fig.~\ref{fig:energydiff_HOIs}(b,c)).
On the other hand, the 4-spin 4-site interaction with spin moments on a tetrahedron is odd (cf.~Eq.~(\ref{eq:H_HOI_perp}) and
Fig.~\ref{fig:energydiff_HOIs}(d,e))
just as the odd 4-spin 3-site term (cf.~Fig.~\ref{fig:BL_HOI}(b)). 

Therefore, the difference in the energy gain of the 3Q state to its respective RW-AFM state in the two alignments originates from the odd interlayer HOI terms. The contribution from even HOI terms moves the 3Q states of Fig.~\ref{fig:energydiff_HOIs}(a) down in energy, in the $\rightrightarrows$ as well as 
the $\rightleftarrows$ alignment of the Mn bilayer.
The odd HOI terms favor the 3Q state in $\rightleftarrows$ alignment even further and move the 3Q$_{\rightrightarrows}$ state up in energy. The contributions from odd and even HOIs are quantitatively different by a factor of approximately two, as seen in Tab.~\ref{tab:eff_contributions},
driving the surprising behavior of energetically disfavored 3Q$_{\rightrightarrows}$ state to its 
RW-AFM$_{\rightrightarrows}$ state and a huge energy gain of the 
3Q$_{\rightleftarrows}$ state to the RW-
AFM$_{\rightleftarrows}$ state of 45~meV. 
This effect is even 
larger than
the energy splitting of the two RW-AFM states due to interlayer exchange interactions.
As a result the ideal bilayer 3Q state with tetrahedron angles between neighboring intra- and interlayer spin moments becomes the ground state of the Mn bilayer on Ir(111) through effects 
of interlayer higher-order interactions. Note, that for different signs of the odd interlayer
HOI terms, e.g.~in another material system,
one could also obtain a situation in which the other 3Q state is stabilized in a bilayer.

\textbf{Generalization to arbitrary multi-Q states.}
Following this classification of HOI terms, we can generalize the result discussed above 
for the
3Q state to an arbitrary multi-Q state and its corresponding 1Q state.
One can write the energy in the Heisenberg model extended by intra- and interlayer
HOI terms of an arbitrary spin state within a layer
and parallel and antiparallel alignment between the two layers as
\begin{equation}
\begin{split}
E^{\rightleftarrows,\rightrightarrows} =  E_{\parallel} 
                \pm  E_{\perp}
                +    E^{\text{HOI}}_{\text{even}}
                \pm  E^{\text{HOI}}_{\text{odd}} 
 \label{eq:eff_energy_contributions}
 \end{split}
\end{equation}
where $E_{\parallel}$ and $E_{\perp}$ correspond to the energy of bilinear intra- and interlayer Heisenberg exchange interactions and $E^{\text{HOI}}_{\text{even}}$ and $E^{\text{HOI}}_{\text{odd}}$ are contributions from even and odd HOIs.
For a multi-Q state (MQ) and its respective 1Q state (1Q) the energy for 
$\rightleftarrows$ and $\rightrightarrows$ alignment follows as 
\begin{equation}
\begin{split}
E^{\rightleftarrows,\rightrightarrows}_{1Q} =
E_{\parallel} 
                \pm  E_{\perp}
                +    E^{\text{1Q}}_{\text{even}}
                \pm  E^{\text{1Q}}_{\text{odd}} \\
E^{\rightleftarrows,\rightrightarrows}_{MQ}=  E_{\parallel} 
                \pm  E_{\perp}
                +    E^{\text{MQ}}_{\text{even}}
                \pm  E^{\text{MQ}}_{\text{odd}} 
 \label{eq:eff_energy1QMQ}
 \end{split}
\end{equation}
indicating the degeneracy in the bilinear exchange ($E_{\parallel}, E_{\perp}$) and the sign flip of odd interactions ($E_{\perp},E^{\text{MQ}}_{\text{odd}}$) regarding the 
$\rightleftarrows$ and $\rightrightarrows$
alignment as discussed above.
Here, the superscript denotes the different HOI contributions to the multi-Q and 1Q-state. 

The effect of the energy contributions to the 1Q and the corresponding multi-Q state 
in the $\rightleftarrows$ and in the
$\rightrightarrows$ alignment as given by Eq.~(\ref{eq:eff_energy_contributions}) 
is illustrated in Fig.~\ref{fig:energydiff_HOIs}(f-i)
for the example of the RW-AFM (1Q) and 3Q states.
If only the interlayer bilinear exchange, $E_{\parallel}$, is considered
(Fig.~\ref{fig:energydiff_HOIs}(f)) 
the 1Q and the 3Q states are energetically degenerate and their energies do
not depend on the spin alignment between the layers.
Upon taking the interlayer bilinear exchange, $E_{\perp}$, into account the
energies for the two spin alignments are shifted with respect to each other
(Fig.~\ref{fig:energydiff_HOIs}(g)).
Going beyond the Heisenberg bilinear exchange to the HOI, we find that the degeneracy
of the 1Q and 3Q states is lifted by the even terms, $E^{\text{HOI}}_{\text{even}}$
(Fig.~\ref{fig:energydiff_HOIs}(h)).
Finally, the contribution from the odd HOI terms, $E^{\text{HOI}}_{\text{odd}}$, 
favors the $\rightleftarrows$ alignment and disfavors the $\rightrightarrows$ alignment for the chosen sign of $E^{\text{HOI}}_{\text{odd}}$
(Fig.~\ref{fig:energydiff_HOIs}(i)). Taking the odd HOI into account -- which
all orginate from interlayer HOI -- allows to
explain the DFT result for the Mn BL on Ir(111) (Fig.~\ref{fig:energydiff_HOIs}(a)).
Note, that we have chosen arbitrary values for
the different energy contributions here to illustrate the effect in a qualitative
way.

We can further derive equations of energy differences, in which only the odd or the even HOI terms 
play a role. Taking the energy difference between the 
$\rightleftarrows$ and $\rightrightarrows$
alignment of one multi-Q state only bilinear interlayer exchange and odd HOI terms remain. Subtracting the same energy difference for the 1Q state leads to 
\begin{equation}
\begin{split}
 (E^{\rightleftarrows}_{MQ} - E^{\rightrightarrows}_{MQ}) - 
 (E^{\rightleftarrows}_{1Q} - E^{\rightrightarrows}_{1Q}) = 
 2( E^{\text{MQ}}_{\text{odd}} - E^{\text{1Q}}_{\text{odd}})
 \label{eq:eff_energy_odd}
 \end{split}
\end{equation}
in which the contribution from bilinear interlayer exchange vanishes and only odd HOI terms are present.

Now by adding the energies of $\rightleftarrows$ and $\rightrightarrows$
aligned layers and subtracting the 1Q state from the multi-Q state, one finds 
\begin{equation}
\begin{split}
   (E^{\rightleftarrows}_{MQ} + E^{\rightrightarrows}_{MQ}) - 
   (E^{\rightleftarrows}_{1Q} + E^{\rightrightarrows}_{1Q}) = 2( E^{\text{MQ}}_{\text{even}} - E^{\text{1Q}}_{\text{even}})
 \label{eq:eff_energy_even}
 \end{split}
\end{equation} 
showing only contributions from even HOIs, due to their indifference regarding the two alignments.

The contributions from odd and even HOI terms can be calculated via Eqs.~(\ref{eq:eff_energy_odd}) and (\ref{eq:eff_energy_even}), respectively,
using the DFT total energies
of the 1Q state and the corresponding multi-Q state. For the Mn bilayer on Ir(111) the
values 
are displayed in Tab.~\ref{tab:eff_contributions}.
For the 3Q state we find that the energy differences of even 
and odd HOI terms have the same sign and the odd HOI terms contribute with about twice
the strength. Subtracting these two values explains the nearly 15 meV energy difference between the 3Q$_\rightrightarrows$ state and the RW-AFM$_\rightrightarrows$ state observed in 
Fig.~\ref{fig:energydiff_HOIs}(a). For the opposite alignment
between the two layers ($\rightleftarrows$) 
the odd and even HOI energy contributions add up which results
in the large energy gain of the 3Q$_\rightleftarrows$ state of about 45 meV with respect to the
RW-AFM$_\rightleftarrows$ state. 
We can apply the approach also for the $uuud$ states along
the two high symmetry directions of the 2D BZ (last two lines
of Tab.~\ref{tab:eff_contributions}).

The contributions from odd and even HOI terms for multi-Q
states in the freestanding Mn BL are given in Tab.~\ref{tab:eff_contributions_2Mn}. Here the contribution at the $\overline{\text{M}'}$-point of the BZ is even larger for the odd HOIs but does not surpass the energy difference of the RW-AFM states in the two alignment. Therefore the RW-AFM$_{\rightrightarrows}$ is still the lowest in energy (as visible at the $\overline{\text{M}'}$-point
in the dispersion Fig.~\ref{fig:Interlayer_MnBL}(a)).
Note, that it is not possible to obtain the seven intra- and interlayer HOI constants from the six energy differences of the two 3Q and
four $uudd$ states,
i.e.~using the system of Eqs.~(\ref{eq:RW-AFM_energy_contributions}),
(\ref{eq:3Q_energy_contributions})
and (\ref{eq:1QGM_energy_contributions}-\ref{eq:UDGK_energy_contributions}).
\begin{table}[htb]
    \centering
    \caption{
    Energy contributions from even and odd HOI terms to the energy difference of 1Q and 
    multi-Q states as obtained via DFT for the Mn bilayer on Ir(111). The three points in 
    the BZ are considered (3Q and RW-AFM state, $uudd$ and 90° spin spiral in $\overline{\Gamma\text{M$'$}}$ and $\overline{\Gamma\text{K$'$}}$ direction). The values 
    are given in meV per two Mn atoms.}
    \begin{ruledtabular}
        \begin{tabular}{lccc}
        $\mathbf{q}$ & states & $E^{\text{MQ}}_{\text{even}} - E^{\text{1Q}}_{\text{even}}$ & $E^{\text{MQ}}_{\text{odd}} - E^{\text{1Q}}_{\text{odd}}$ \\
        \hline
         $\overline{\text{M$'$}}$ &  3Q/RW-AFM & $-15.46$ & $-31.08$ \\ %
         $\frac{1}{2}\overline{\Gamma\text{M$'$}}$ &  uudd/90°-spiral & $10.67$ & $6.67$\\ %
         $\frac{3}{4}\overline{\Gamma\text{K$'$}}$ &  uudd/90°-spiral &  $-7.64$ &  $18.08$ %
          \end{tabular}
    \label{tab:eff_contributions}
    \end{ruledtabular}
\end{table}

Recently, the 3Q state  was also 
discovered in the bulk crystal Co$_{1/3}$TaS$_2$, in which 
hexagonal layers of Co atoms
are separated by van der Waals spacer layers   
\cite{park2023tetrahedral,takagi2023spontaneous}.
Based on neutron scattering and transport measurements it was concluded that
a 3Q state exists within each of the Co layers with an
$\rightleftarrows$ alignment
between layers
as discussed in our work for Mn bilayers.
Ref.~\cite{park2023tetrahedral} discussed their findings also through a classical spin model containing intralayer Heisenberg and higher-order exchange 
($J_1^\parallel$, $J_2^\parallel$, $B_1^\parallel$) for  realizing the 3Q spin structure within the layers and a NN interlayer exchange ($J_1^\perp<0$) aligns the layers in the net antiferromagnetic alignment of the ideal bilayer 3Q state.
However, this spin model only takes contributions of $E_{\parallel}, E_{\perp}, E_{\mathrm{even}}^{\mathrm{HOI}}$ from Eq.~(\ref{eq:eff_energy_contributions}) into account, shown qualitatively in Fig.~\ref{fig:energydiff_HOIs}(f-h)
and does not consider interlayer HOI. 

In contrast, we find that 
the stabilization of the 
3Q$_\rightleftarrows$ state (Fig.~\ref{fig:BL_magnetisation}(c))
in Mn bilayers on Ir(111) 
(Fig.~\ref{fig:energydiff_HOIs}(a)) cannot be
explained by intralayer HOI and interlayer Heisenberg exchange. We show
that interlayer HOI acting between spins in the 
two layers must be taken into account, 
specifically $E_{\mathrm{odd}}^{\mathrm{HOI}}$ (Fig.~\ref{fig:energydiff_HOIs}(i)). 
We introduce interlayer higher-order interactions in a NN framework and 
in a general way, which were to our knowledge not studied previously. 
These terms can favor either the bilayer 3Q state 
3Q$_{\rightleftarrows}$ (Fig.~\ref{fig:BL_magnetisation}(c))
or the net ferromagnetically aligned 3Q state 3Q$_{\rightrightarrows}$ 
(Fig.~\ref{fig:BL_magnetisation}(d))
depending on the sign of the interlayer HOI.

\section{Summary and Conclusion}
\label{sec:Conclusion}

We have studied Mn mono- and bilayers on the 
Ir(111) surface using
first-principles calculations based on DFT. 
For the Mn monolayer,
the magnetic ground state depends on the stacking sequence.
The energy dispersion
of spin spirals exhibits the lowest state for both fcc and hcp stacking
at the $\overline{\rm K}$ point of the BZ corresponding to the 
N\'eel state with 120$^\circ$ angles between nearest neighbor spin
moments. However, for fcc stacking the 3Q state with tetrahedron
angles between nearest neighbors is energetically even lower. The
stabilization of the 3Q state is a result of the energy gain due
to intralayer higher-order exchange interactions. 

For the Mn bilayer, we demonstrate based on spin spiral calculations 
that the antiferromagnetic pair-wise intra- and interlayer exchange interactions favor a row-wise antiferromagnetic (RW-AFM) state with a net antiferromagnetic alignment ($\rightleftarrows$)
between spin moments of the Mn layers. Unexpectedly, the energy gain of the 3Q state due to higher-order interactions is much larger than the energy splitting between the two row-wise antiferromagnetic states with opposite interlayer couplings. 
In the predicted ideal bilayer 3Q state of the Mn bilayer on Ir(111) the 
magnetic moments exhibit tetrahedron angles between all nearest neighbor sites, i.e.~within a magnetic layer and between the two magnetic layers. Thereby, the topological orbital moments of
the two layers -- which emerge due to the non-coplanarity of the spin structure -- 
point into the same direction leading to a significant total orbital magnetization \cite{Saxena2024}.

In order to explain the surprising observations for the Mn
bilayer from our DFT calculations
we have developed an atomistic spin model for magnetic bilayers
including intra- and interlayer higher-order exchange interactions. We show that the higher-order interactions can be classified into even and 
odd terms with respect to flipping the spin moments in one of the magnetic layers. This allows to distinguish quite generally the effect of HOI 
terms on the energy difference between a multi-Q state and its corresponding 1Q state.
Based on our spin model, we explain that the bilayer 3Q state in a 
Mn bilayer on Ir(111) is stabilized by interlayer higher-order 
exchange interactions. The large
energy gain of the bilayer 3Q state with tetrahedron angles between all nearest neighbors obtained from DFT
is explained by an additive effect of even and odd higher-order exchange contributions.

\section*{Acknowledgments} 
It is our pleasure to thank Kirsten von Bergmann, Vishesh Saxena, and Andr\'e Kubetzka for valuable discussions. 
We gratefully acknowledge financial support from the Deutsche Forschungsgemeinschaft (DFG, German Research Foundation) via SPP2137 “Skyrmionics” (project no.~462602351),  project no.~418425860 
and project no.~555842692.
The authors gratefully acknowledge the computing time made available to them on the high-performance computers ``Lise" at the NHR center NHR@ZIB and ``Emmy" at the NHR center NHR@G\"{o}ttingen. This center is jointly supported by the Federal Ministry of Education and Research and the state governments participating in the NHR~\cite{nhr}.

\appendix
\section{Fitting functions for energy dispersions of magnetic mono- and bilayers}
\label{sec:fittingfunctions}

\begin{figure}[htb]
	\centering
\includegraphics[width=0.85\linewidth]{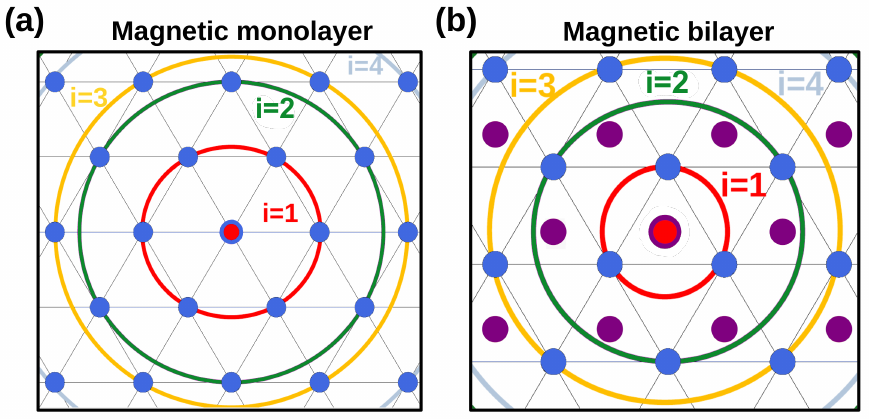}
	\caption{Illustration of circular rings 
    showing equidistant neighbors (shells) up to the 3rd nearest neighbor 
    for a reference atom on the hexagonal lattice in real space  of (a) a magnetic monolayer and (b) a magnetic bilayer. Summing up the relative atomic positions of the $\sigma$th shell via Eq.~(\ref{eq:scalarprodukt_spins}) and the first term of  Eq.~(\ref{eq:H_general}) yields an analytical expression for the pairwise (a) intralayer and (b) interlayer Heisenberg exchange which is further utilized to fit the respective DFT calculated energy dispersion.}
	\label{fig:fittingfunctions}
\end{figure}

In this section we briefly explain how to calculate the fitting functions for the parameters of the pairwise Heisenberg exchange on the atomic hexagonal lattice of both a magnetic mono- and bilayer. Since our investigation focuses on periodic magnetic structures, it is convenient to represent the localized magnetic moments of the Heisenberg model by their discrete Fourier components~\cite{Kurz2000}  
\begin{equation}
 \mathbf{M}_i=\sum \limits_{\mathbf{q}}\mathbf{M}_{\mathbf{q}}e^{i\mathbf{q}\mathbf{R}_i}\textbf{,}
\end{equation}
with the sum running over the reciprocal lattice vectors $\mathbf{q}$ and the real space coordinates $\mathbf{R}_i$. By substituting the localized spins with their Fourier components in the Hamiltonian for the pairwise Heisenberg exchange (see first term of Eq. (\ref{eq:H_general})) and transforming the resulting expression further, we obtain the Heisenberg exchange interaction on the reciprocal lattice in terms of the Fourier transforms of the exchange constants $J(\mathbf{q})$: 
\begin{equation}
    E_{{\text{exc}}} (\mathbf{q})=-N M^2 J(\mathbf{q}) =-N M^2 \sum\limits_\sigma J_\sigma \sum\limits_{\mathbf{r} \in \sigma} e^{-i\mathbf{q}\mathbf{r}} 
\end{equation}
Here, $M$ is the magnetic moment per site, $N$ gives the number
of lattice sites, and the index $\sigma$ denotes the number of shells taken into account for the summation of interacting magnetic moments \textcolor{blue}{(cf.~Fig.~\ref{fig:fittingfunctions})}. As the exchange interaction is isotropic, spins with the same distance from a reference atom on the hexagonal lattice interact with the same exchange constant and can hence be grouped into one shell which is illustrated in Fig.~\ref{fig:fittingfunctions}(a). 
The fundamental solutions of the classical Heisenberg model on a periodic Bravais lattice can be written as~\cite{Kurz2000}
\begin{equation}
 \mathbf{M}_i=M \Bigl( 
 \mathbf{e}_1
 \cos(\mathbf{q}\cdot\mathbf{R}_i)+
 \mathbf{e}_2
 \sin(\mathbf{q}\cdot\mathbf{R}_i) \Bigr).
 \label{eq:magmom_helical}
\end{equation}
This expression denotes a spin spiral in which the orthogonal
unit vectors $\mathbf{e}_1$ and $\mathbf{e}_2$ span the rotational
plane. In the absence of spin-orbit coupling the vectors can be
assumed to span the $xy-$plane without loss of generality.
Consequently, the resulting magnetic texture describes a flat spin spiral state in which the magnetic moments rotate by a constant angle from lattice site to lattice site around the $z$ axis along the propagation direction given by the spin spiral vector $\mathbf{q}$. Rewriting the dot product between two spins as it appears in the Hamiltonian of the pairwise Heisenberg exchange (first term of Eq. (\ref{eq:H_general})) via Eq.~(\ref{eq:magmom_helical}), one obtains the following term:  
\begin{equation}
 \mathbf{M}_i \cdot \mathbf{M}_j
 =M^2 \cos(\mathbf{q}\cdot (\mathbf{R}_i - \mathbf{R}_j))
 \label{eq:scalarprodukt_spins}
\end{equation}
This simplified expression can now be used to compute a fitting function including a desired number of nearest neighbor shells for the DFT calculated energy dispersion of a magnetic monolayer. With the two-dimensional spin spiral vector taking the form  
\begin{equation}
 \mathbf{q}=\frac{2\pi}{a}(q_x,q_y)^\text{T}\text{,}
\end{equation}
the evaluation of the sum of the atomic positions of eight shells with respect to the position of a reference atom yields the analytical form of the pairwise intralayer Heisenberg exchange on the hexagonal lattice:
\begin{widetext}
\begin{align}
 E_{\text{exc}}=& - J_1^{\parallel} \Bigl[2\cos(2\pi q_x)+4\cos(\pi q_x)\cos(\pi \sqrt{3}q_y) \Bigr] \nonumber \\
 & - J_2^{\parallel} \Bigl[2\cos(2\pi \sqrt{3} q_y)+4\cos(3\pi q_x)\cos(\pi \sqrt{3}q_y) \Bigr] \nonumber \\
 & - J_3^{\parallel} \Bigl[2\cos(4 \pi q_x)+4\cos(2\pi q_x)\cos(2\pi \sqrt{3}q_y) \Bigr] \nonumber \\
 & - J_4^{\parallel} \Bigl[4\cos(\pi q_x)\cos(3\pi \sqrt{3}q_y)+4\cos(4\pi q_x)\cos(2\pi \sqrt{3}q_y) + 4\cos(5\pi q_x)\cos(\pi \sqrt{3}q_y) \Bigr]\nonumber \\
 & - J_5^{\parallel} \Bigl[2\cos(6 \pi q_x)+4\cos(3\pi q_x)\cos(3\pi \sqrt{3}q_y) \Bigr] \nonumber \\
 & - J_6^{\parallel} \Bigl[2\cos(4 \pi \sqrt{3}q_y)+4\cos(6\pi q_x)\cos(2\pi \sqrt{3}q_y) \Bigr] \nonumber \\
 & - J_7^{\parallel} \Bigl[4\cos(2\pi q_x)\cos(4\pi \sqrt{3}q_y)+4\cos(5\pi q_x)\cos(3\pi \sqrt{3}q_y) + 4\cos(7\pi q_x)\cos(\pi \sqrt{3}q_y) \Bigr]\nonumber \\
 & - J_8^{\parallel} \Bigl[2\cos(8 \pi q_x)+4\cos(4\pi q_x)\cos(4\pi \sqrt{3}q_y) \Bigr] \nonumber \\
 \label{eq:Intralayer_exchange}
\end{align}
\end{widetext}
Note that in principle any number of shells can be considered for the fitting function, however, since the influence of the pairwise exchange quickly decreases upon increasing the distance between two magnetic moments, we restrict ourselves to the inclusion of eight shells
of neighboring atomic sites. 

\begin{figure*}[htb]
	\centering
	\includegraphics[width=0.6\textwidth]{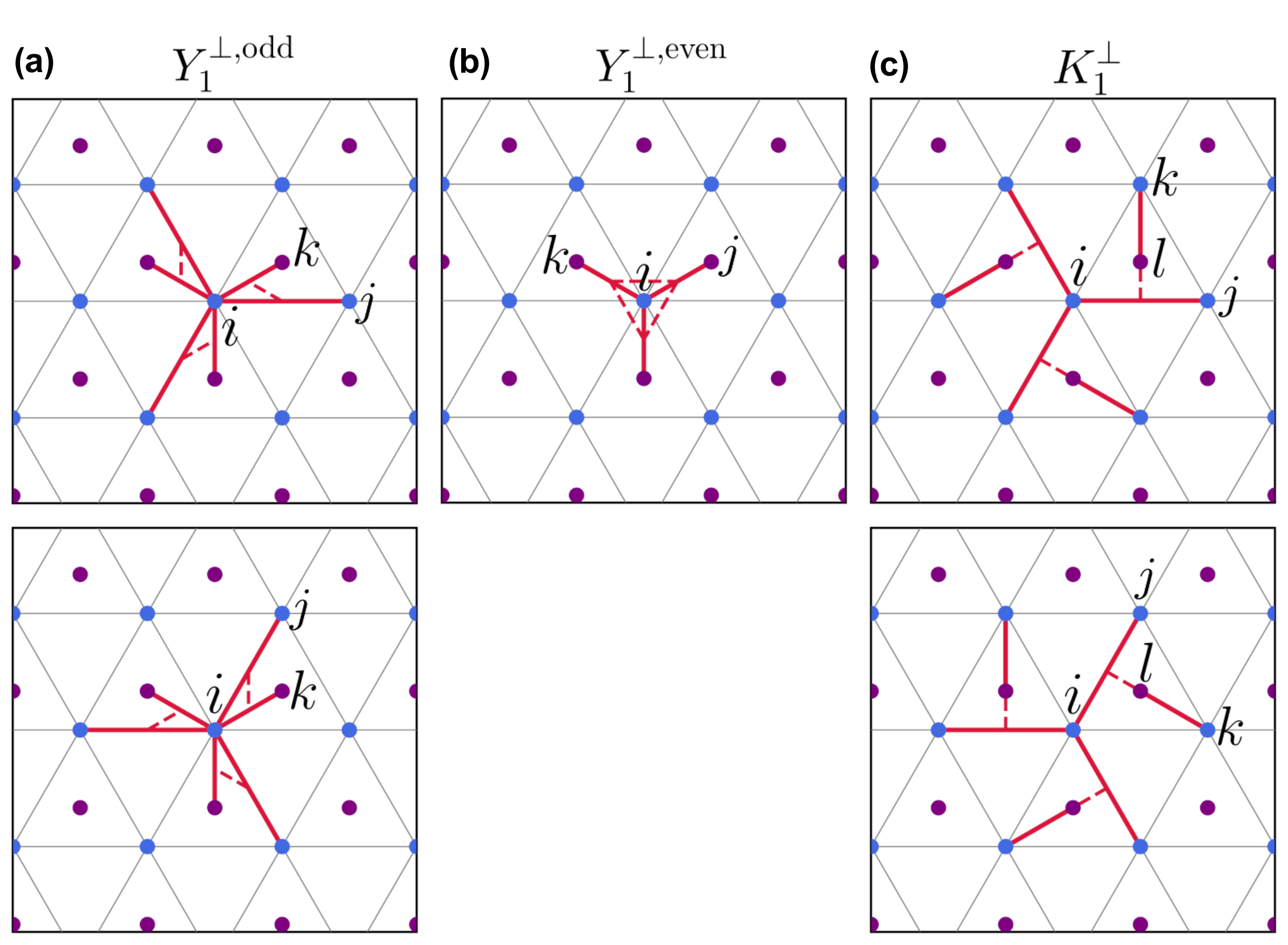}
	\caption{
     Sketch of a hexagonal bilayer showing the pairs of atomic sites contributing to the interlayer higher-order interactions. 
     (a) The odd interlayer 4-spin 3-site interaction, $Y_1^{\perp,\text{odd}}$. 
     (b) The even interlayer 4-spin 3-site interaction, $Y_1^{\perp,\text{even}}$. 
     (c) The odd interlayer 4-spin 4-site interaction, $K_1^{\perp}$.
     The atoms of the top layer (purple dots) lie in the hollow site of the bottom layer (blue dots). The four atomic sites participating in an interaction are connected by red lines, the scalar products (solid lines) are connected by a dotted line of the same color. Every panel displays the pairs given from the summation with a fixed atom $i$ of one of the higher-order interactions. Every panel shows three pairs each. In (a,c) the pairs of the upper and lower panels are transferable through a C3-symmetry, adding up to six pairs for (a,c) and three for the $Y_1^{\perp,\text{even}}$ interaction (b). For the three-site terms (a,b) the indices $i,j,k$ are shown
     for one pair of spins and for the four-site term (c) the indices $i,j,k,l$ are given for one term.
     }
\label{fig:HOI_pairs_interlayer}
\end{figure*}

Applying the same approach to a magnetic bilayer, the fitting function for the pairwise interlayer Heisenberg exchange for one reference spin in the top magnetic layer interacting with neighboring spins up to the eighth order in the lower layer (the same applies vice versa) is obtained as:
\begin{widetext}
\begin{align}
  E_{\text{exc}} =& -J_1^{\perp}\Bigl[\cos(2\pi\frac{\sqrt{3}}{3}q_y)+2\cos(\pi q_x)\cos(\pi \frac{\sqrt{3}}{3}q_y)\Bigr] \nonumber \\
 &-J_2^{\perp}\Bigl[2\cos(2\pi q_x)\cos(2\pi\frac{\sqrt{3}}{3}q_y) + \cos(4\pi\frac{\sqrt{3}}{3}q_y)\Bigr] \nonumber \\
 &-J_3^{\perp}\Bigl[2\cos(\pi q_x)\cos(5\pi\frac{\sqrt{3}}{3}q_y)+2\cos(3\pi q_x)\cos(\pi\frac{\sqrt{3}}{3}q_y)+2\cos(2\pi q_x)\cos(4\pi\frac{\sqrt{3}}{3}q_y)\Bigr] \nonumber \\
 &-J_4^{\perp}\Bigl[2\cos(3\pi q_x)\cos(5\pi\frac{\sqrt{3}}{3}q_y)+2\cos(4\pi q_x)\cos(2\pi\frac{\sqrt{3}}{3}q_y)+2\cos(\pi q_x)\cos(7\pi\frac{\sqrt{3}}{3}q_y)\Bigr] \nonumber \\
 &-J_5^{\perp}\Bigl[\cos(8\pi\frac{\sqrt{3}}{3}q_y)+2\cos(4\pi q_x)\cos(4\pi\frac{\sqrt{3}}{3}q_y)\Bigr] \nonumber \\
 &-J_6^{\perp}\Bigl[2\cos(2\pi q_x)\cos(8\pi\frac{\sqrt{3}}{3}q_y)+2\cos(5\pi q_x)\cos(\pi\frac{\sqrt{3}}{3}q_y)+2\cos(3\pi q_x)\cos(7\pi\frac{\sqrt{3}}{3}q_y)\Bigr] \nonumber \\
 &-J_7^{\perp}\Bigl[\cos(10\pi\frac{\sqrt{3}}{3}q_y)+2\cos(5\pi q_x)\cos(5\pi\frac{\sqrt{3}}{3}q_y)\Bigr] \nonumber \\
 &-J_8^{\perp}\Bigl[2\cos(4\pi q_x)\cos(8\pi\frac{\sqrt{3}}{3}q_y)+2\cos(2\pi q_x)\cos(10\pi\frac{\sqrt{3}}{3}q_y)+2\cos(6\pi q_x)\cos(2\pi\frac{\sqrt{3}}{3}q_y)\Bigr] \nonumber \\
\end{align}
\end{widetext}
Note that the interlayer exchange indeed requires a new analytical function since the shift of the two atomic sublattices of the magnetic bilayer results in other distances between interacting neighboring spins compared to the monolayer. As shown in Fig.~\ref{fig:fittingfunctions}(b), for a central reference atom located in one sublattice only three nearest and next-nearest neighbors in the other magnetic layer are left due to the lost rotational symmetry of the hexagonal lattice around the $z$ axis. 
\section{Exchange constants for the Mn ML on Ir(111)}
In the main part of the paper a detailed discussion of the energy dispersion, the derived intralayer Heisenberg exchange parameters, the higher-order terms (HOI) as well as the influence of the Dzyaloshinskii-Moriya interaction (DMI) on the spin spiral states has been presented for both stackings of the Mn ML on Ir(111). For the sake of completeness, the full set of exchange constants obtained from DFT is listed in Tables~\ref{tab:Heisenberg_exchange_MnIr111},~\ref{tab:DMI_constants_MnIr111} and~\ref{tab:HOI_constants_MnIr111}.   
\begin{table*} [!htbp]
	\centering
    \caption{Intralayer Heisenberg exchange constants ($J_{i}^{\parallel}$) calculated via DFT for fcc- and hcp-stacked Mn MLs on Ir(111). The values have been extracted from fitting the spin spiral energy dispersion $E(\mathbf{q})$ neglecting HOI shown in Fig.~\ref{fig:MnIr111_dispersion}(a) of the main text. $J_i^{\parallel}>0$ ($J_i^{\parallel}<0$) denotes ferromagnetic (antiferromagnetic) coupling.
    All values are given in meV/Mn atom.
		}
	\begin{ruledtabular}
		\begin{tabular}{lcccccccc}
			$\mathrm{System}$ & $J_{1}^{\parallel}$ & $J_{2}^{\parallel}$ & $J_{3}^{\parallel}$ & $J_{4}^{\parallel}$ & $J_{5}^{\parallel}$ & $J_{6}^{\parallel}$ & $J_{7^{\parallel}}$ & $J_{8}^{\parallel}$  \\
			\colrule
			fcc-Mn/Ir(111) 
   & $-31.22$ & $-$4.29 & $-1.72$ & $-0.30$ & 0.00 & $-0.19$ & 0.08& 0.01 \\
			hcp-Mn/Ir(111) 
   & $-36.78$ & $-4.47$ & $-1.02$ & $-0.18$ &0.01 &0.01& 0.08& $-0.10$ \\
		\end{tabular}
		\label{tab:Heisenberg_exchange_MnIr111}
	\end{ruledtabular}
\end{table*}

\begin{table*}[htb]
	\centering
	\caption{Intralayer Dzyaloshinskii-Moriya interaction constants ($D_i^{\parallel}$) and magnetocrystalline anisotropy energy constant ($K_{u}$) for fcc- and hcp-stacked Mn MLs on Ir(111) calculated via DFT. $D_i>0$ ($D_i<0$) denotes a preference of clockwise (counterclockwise) rotation of spin spirals and a positive  value of $K_{u}$ (obtained for the RW-AFM state) an in-plane easy magnetization axis. All values are given in meV/Mn atom.}
	\begin{ruledtabular}
		\begin{tabular}{l c c c c c c}
			System & $D_1^{\parallel}$ & $D_2^{\parallel}$ &$D_3^{\parallel}$ & $D_4^{\parallel}$ & $D_5^{\parallel}$ & $K_{u}$   \\ 
			\colrule
			fcc-Mn/Ir(111) 
   & 2.77  & 0.26  & 0.27 & $-$0.00 &0.04 & 1.44\\
			hcp-Mn/Ir(111) 
   & 2.82  & 0.44  & 0.28 & $-0.02$ & $-0.04$&1.72 \\
			 \end{tabular}
    \label{tab:DMI_constants_MnIr111}
	\end{ruledtabular}
\end{table*}

\begin{table*}[htb]
	\centering
	\caption{DFT calculated intralayer higher-order exchange interactions for both stackings of the Mn ML on Ir(111). 4-spin 4-site ($K_1^{\parallel}$), biquadratic ($B_1^{\parallel}$) and 4-spin 3-site interaction ($Y_1^{\parallel}$) constants calculated from Eqs. (\ref{eq:HOI_energy1})-(\ref{eq:HOI_energy3}). $\Delta E$ terms the energy difference between the multi-Q and the corresponding 1Q spin spiral state, respectively, i.e., $\Delta E_{\frac{1}{2}\overline{\Gamma\text{M}}}= E^{\text{uudd}}_{\frac{1}{2}\overline{\Gamma\text{M}}} - E_{\frac{1}{2}\overline{\Gamma\text{M}}}^{\text{1Q}}$, $\Delta E_{\frac{3}{4}\overline{\Gamma\text{K}}}= E^{\text{uudd}}_{\frac{3}{4}\overline{\Gamma\text{K}}} - E_{\frac{3}{4}\overline{\Gamma\text{K}}}^{\text{1Q}}$ and $\Delta E_{\overline{\text{M}}} = E_{\overline{\text{M}}}^{\text{3Q}}-E_{\overline{\text{M}}}^{\text{1Q}}$.  All values are given in meV/Mn atom.}
	\begin{ruledtabular}
		\begin{tabular}{l c c c c c c}
			System & $B_1^{\parallel}$ & $Y_1^{\parallel}$ &$K_1^{\parallel}$ & $\Delta E_{\frac{1}{2}\overline{\Gamma\text{M}}}$ & $\Delta E_{\frac{3}{4}\overline{\Gamma\text{K}}}$ & $\Delta E_{\overline{\text{M}}}$   \\ 
			\colrule
			fcc-Mn/Ir(111) 
   & $-3.19$  & $-3.02$  & $-1.22$ & $+$15.05 &$-9.08$ & $-13.95$\\
			hcp-Mn/Ir(111) 
   & $-3.19$  & $-3.50$  & $-0.89$ & $+19.60$ & $-8.36$&$-7.87$ \\
			 \end{tabular}
    \label{tab:HOI_constants_MnIr111}
	\end{ruledtabular}
\end{table*}

\section{DMI contribution to the spin spiral dispersion 
for the Mn BL on Ir(111)}
\label{appendix:DMI}
\begin{figure*}[htb]
	\centering
	\includegraphics[width=0.82\textwidth]{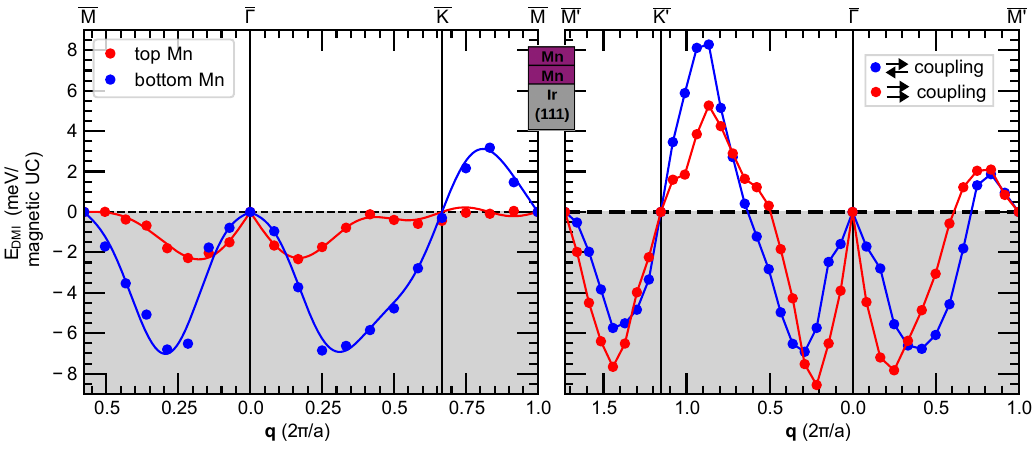}
	\caption{Contributions of the Dzyaloshinskii-Moriya interaction (DMI) for cycloidal spin spiral states in the Mn BL on Ir(111). (a) Intralayer DMI obtained for $\mathbf{q}$ values of spin spirals in the 2D hexagonal BZ which propagate in only one magnetic layer, i.e. either the top or bottom Mn layer (the corresponding scalar-relativistically calculated energy dispersions  are shown in Fig.~\ref{fig:Intralayer_MnBL}(b)). The solid lines running through the DFT values denote a fit to the intralayer DMI comprising five nearest-neighbor shells. (b) DMI contributions for all $\mathbf{q}$ values of spin spirals propagating simultaneously in both magnetic layers (the respective scalar-relativistically calculated energy dispersion for a $\rightleftarrows$ or $\rightrightarrows$ alignment of the two layers at the 
    $\overline{\text{M}'}$-point
    is displayed in Fig.~\ref{fig:Interlayer_MnBL}(b)). Here, the solid lines connecting the DFT data points serve as a guide to the eye. $E_{\text{DMI}}<0 (>0)$: preference of right-(left-)rotating cycloidal spin spirals. All values are given in meV/magnetic u.c., i.e. per two Mn atoms. } 
	\label{fig:DMI_MnBLIr111}
\end{figure*}

In Fig.~\ref{fig:DMI_MnBLIr111} we show the energy contributions of the DMI for the different types of spin spiral setups in the Mn BL on Ir(111) (cf.~Fig.~\ref{fig:Intralayer_MnBL}(b) and Fig.~\ref{fig:Interlayer_MnBL}(b) for the respective scalar-relativistic energy dispersions). Note that the maximum deviation for the DMI between the $\rightleftarrows$ and 
the $\rightrightarrows$ alignment %
of the magnetic layers in Fig.~\ref{fig:DMI_MnBLIr111}(b) amounts to only 4 meV/magnetic unit cell indicating the interlayer SOC effect to play a minor role for this ultrathin film system.

\begin{table*} [htbp]
	\centering
    \caption{DFT calculated energy differences between multi- and single-Q states for a freestanding Mn bilayer and the hcp-Mn/hcp-Mn/Ir(111) film system. The values $\Delta E$ are calculated explicitly as follows:
    $\Delta E_{\frac{1}{2}\overline{\Gamma\text{M$'$}}}^{\rightleftarrows , \rightrightarrows}= E_{\frac{1}{2}\overline{\Gamma\text{M$'$}},\text{uudd}}^{\rightleftarrows , \rightrightarrows} - E_{\frac{1}{2}\overline{\Gamma\text{M$'$}},\text{1Q}}^{\rightleftarrows , \rightrightarrows}$,
    $\Delta E_{\frac{3}{4}\overline{\Gamma\text{K$'$}}}^{\rightleftarrows , \rightrightarrows}= E_{\frac{3}{4}\overline{\Gamma\text{K$'$}},\text{uudd}}^{\rightleftarrows , \rightrightarrows} - E_{\frac{3}{4}\overline{\Gamma\text{K$'$}},\text{1Q}}^{\rightleftarrows , \rightrightarrows}$
    and
    $\Delta E_{\text{3Q}}^{\rightleftarrows , \rightrightarrows} = E_{\text{3Q}}^{\rightleftarrows , \rightrightarrows}-E_{\text{RW-AFM}}^{\rightleftarrows , \rightrightarrows}$.
    All values are given in meV/per two Mn atoms in order to be able to directly compare them to the reference 1Q states. Note that $\rightleftarrows , \rightrightarrows$ indicates the alignment of the Mn layers as described in Fig.~\ref{fig:BL_magnetisation}.}
	\begin{ruledtabular}
		\begin{tabular}{lcccccc}
			$\mathrm{System}$ & $\Delta E_{\text{3Q}}^{\rightleftarrows}$ & $\Delta E_{\text{3Q}}^{\rightrightarrows}$ & $\Delta E_{\frac{3}{4}\overline{\Gamma\text{K$'$}}}^{\rightleftarrows}$ & $\Delta E_{\frac{3}{4}\overline{\Gamma\text{K$'$}}}^{\rightrightarrows}$ &$\Delta E_{\frac{1}{2}\overline{\Gamma\text{M$'$}}}^{\rightleftarrows}$ & $\Delta E_{\frac{1}{2}\overline{\Gamma\text{M$'$}}}^{\rightrightarrows}$ \\
			\colrule
			  Mn/Mn 
   & $-39.20$ & 65.70 & 82.72 & 26.12 &50.10 &$-4.38$ \\
           hcp-Mn/hcp-Mn/Ir(111)
           & $-46.54$ %
           & $15.62$ %
           & 10.44 & $-25.72$ &17.34 &$-4.00$\\
		\end{tabular}
		\label{tab:MultiQ_Ediff_bilayer}
	\end{ruledtabular}
\end{table*}

\begin{table}[htb]
    \centering
    \caption{
        Energy contributions from even and odd HOI terms to the energy difference of 1Q and 
        multi-Q states obtained via DFT for the free-standing Mn bilayer through Eqs.~(\ref{eq:eff_energy_odd}) and (\ref{eq:eff_energy_even}) . The three points in 
        the BZ are considered ($3Q$ and RW-AFM state, $uudd$ and 90° spin spiral in $\overline{\Gamma\text{M$'$}}$ and $\overline{\Gamma\text{K$'$}}$ direction). The values 
        are given in meV per two Mn atoms.
    }
    \begin{ruledtabular}
        \begin{tabular}{lccc}
        $\mathbf{q}$ & states & $E^{\text{MQ}}_{\text{even}} - E^{\text{1Q}}_{\text{even}}$ & $E^{\text{MQ}}_{\text{odd}} - E^{\text{1Q}}_{\text{odd}}$ \\
        \hline
         $\overline{\text{M$'$}}$ &  3Q/RW-AFM & $13.25$ & $-52.45$ \\ 
         $\frac{1}{2}\overline{\Gamma\text{M$'$}}$ &  uudd/90°-spiral & $27.24$ & $22.86$\\ 
         $\frac{3}{4}\overline{\Gamma\text{K$'$}}$ &  uudd/90°-spiral & $54.42$ & $28.30$
          \end{tabular}
    \label{tab:eff_contributions_2Mn}
    \end{ruledtabular}
\end{table}

\section{Analytical expressions for the energy of 90° spin spiral and uudd-states in bilayers}\label{app:E_uudd_90deg}
The main text focuses on the RW-AFM and the 3Q state, which arise from a spin spiral vector $\mathbf q$ at the $\overline{\text{M}'}$-point of the BZ and the energy contributions of the Hamiltonian are discussed. Here (Eqs.~\ref{eq:1QGM_energy_contributions},\ref{eq:UDGM_energy_contributions},\ref{eq:1QGK_energy_contributions},\ref{eq:UDGK_energy_contributions}) the energy contributions of
the symmetrical bilayer Hamiltonian including HOIs $H = H_{\text{s-BL}} + H_{\text{HOI}}^{\parallel} + H_{\text{HOI}}^{\perp}$
to the 90°-spin spirals and $uudd$ states are shown.
$\rightleftarrows$, $\rightrightarrows$ denote the alignment of the bilayer magnetization according to Fig.~\ref{fig:BL_magnetisation}.

\begin{widetext}
\begin{equation}
\begin{split}
E_{\frac{1}{2}\overline{\Gamma\text{M$'$}},\text{1Q}}^{\rightleftarrows , \rightrightarrows} 
=                   & 2 J_1^{\parallel}
                    - 2 B_1^{\parallel}  
                    + 0 Y_1^{\parallel}
                    - 12 K_1^{\parallel}
                    \mp 1 J^{\perp}_1
                    - 1 B^{\text{int}}_1                  
                    + 0 Y_1^{\perp,\text{odd}}
                    + 0 Y_1^{\perp,\text{even}}
                    \pm 2 K_1^{\perp}
 \label{eq:1QGM_energy_contributions}
 \end{split}
\end{equation}

\begin{equation}
\begin{split}
E_{\frac{1}{2}\overline{\Gamma\text{M$'$}},\text{uudd}}^{\rightleftarrows , \rightrightarrows} 
=                   & 2 J_1^{\parallel}
                    - 6 B_1^{\parallel}  
                    + 4 Y_1^{\parallel}
                    - 4 K_1^{\parallel}
                    \mp 1 J^{\perp}_1
                    - 3 B^{\text{int}}_1                  
                    \mp 2 Y_1^{\perp,\text{odd}}
                    + 2 Y_1^{\perp,\text{even}}
                    \pm 6 K_1^{\perp}
 \label{eq:UDGM_energy_contributions}
 \end{split}
\end{equation}

\begin{equation}
\begin{split}
E_{\frac{3}{4}\overline{\Gamma\text{K$'$}},\text{1Q}}^{\rightleftarrows , \rightrightarrows} 
=                   & - 2 J_1^{\parallel}
                    - 2 B_1^{\parallel}  
                    + 0 Y_1^{\parallel}
                    - 12 K_1^{\parallel}
                    \pm 1 J^{\perp}_1
                    - 1 B^{\text{int}}_1                  
                    + 0 Y_1^{\perp,\text{odd}}
                    + 0 Y_1^{\perp,\text{even}}
                    \pm 2 K_1^{\perp}
 \label{eq:1QGK_energy_contributions}
 \end{split}
\end{equation}

\begin{equation}
\begin{split}
E_{\frac{3}{4}\overline{\Gamma\text{K$'$}},\text{uudd}}^{\rightleftarrows , \rightrightarrows} 
=                   & - 2 J_1^{\parallel}
                    - 6 B_1^{\parallel}  
                    - 4 Y_1^{\parallel}
                    - 4 K_1^{\parallel}
                    \pm 1 J^{\perp}_1
                    - 3 B^{\text{int}}_1                  
                    \pm 2 Y_1^{\perp,\text{odd}}
                    - 2 Y_1^{\perp,\text{even}}
                    \pm 6 K_1^{\perp}
 \label{eq:UDGK_energy_contributions}
 \end{split}
\end{equation}    
\end{widetext}

%\bibliography{Reference}

\end{document}